\documentstyle[12pt,psfig,rotating]{article}

\renewcommand{\theequation}{\arabic{section}.\arabic{equation}}
\newcommand{\lbl}[1]{\label{eq:#1}}
\newcommand{ \rf}[1]{(\ref{eq:#1})}
\newcommand{\vs}[1]{\rule[- #1 mm]{0mm}{#1 mm}}
\newskip\humongous \humongous=0pt plus 1000pt minus 1000pt

\newif\ifdtup

\def\mhat{{\widehat m}}

\newcommand{\eq}{\vs{2}\begin{equation}}
\newcommand{\en}{\\[2mm]\end{equation}}
\newcommand{\bea}{\begin{eqnarray}}
\newcommand{\ena}{\end{eqnarray}}
\newcommand{\lapprox}{%
\mathrel{%
\setbox0=\hbox{$<$}
\raise0.6ex\copy0\kern-\wd0
\lower0.65ex\hbox{$\sim$}
}}
\newcommand{\gapprox}{%
\mathrel{%
\setbox0=\hbox{$>$}
\raise0.6ex\copy0\kern-\wd0
\lower0.65ex\hbox{$\sim$}
}}

\newcommand{\cp}[1]{[#1]_+}
\newcommand{\cm}[1]{[#1]_-}
\newcommand{\al}{\alpha}
\newcommand{\be}{\beta}
\newcommand{\ga}{\gamma}

\newcommand{\NP}[1]{Nucl.\ Phys.\ {\bf #1}}

\begin{document}
\renewcommand{\thefootnote}{\fnsymbol{footnote}}

\setcounter{equation}{0}
\setcounter{subsection}{0}
\setcounter{table}{0}
\setcounter{figure}{0}

\begin{titlepage}
\begin{flushright}
CPT-98/P.3716\\
LU TP 98/23\\
ZU-TH 21/98\\
\today\\
\end{flushright}
\begin{center}
\begin{bf}
{\Large \bf Low--energy photon--photon fusion into three pions \\[0.5cm]
in Generalized Chiral Perturbation
Theory  \footnote{Work supported in part by TMR, EC--Contract No. 
ERBFMRX--CT980169} } \\[2cm]
\end{bf}
{\large Ll. Ametller$^a$, J. Kambor$^b$, M. Knecht$^c$ and P. Talavera$^{d}$}
\\[1cm]
$^a$ Dept. de F{\'\i}sica i Enginyeria Nuclear,
UPC, E-$08034$ Barcelona, Spain.\\[.1cm]
$^b$   Institut f\"ur Theoretische Physik,
Universit\"at Z\"urich, 
CH-8057 Z\"urich, Switzerland.\\[.1cm]
$^c$ Centre de physique Th\'eorique, CNRS--Luminy, 
Case 907
F--13288 Marseille Cedex 9, France.\\[.1cm]
$^d$ Dept. of Theoretical Physics, University of Lund, S\"olvegatan 14A,
S-22362 Lund, Sweden. \\[3cm]
{\bf PACS:}~11.30.Rd, 12.39.Fe, 13.75.-n, 14.70.Bh, 13.60.Le\\[0.2mm]
{\bf Keywords:} \begin{minipage}[t]{9.5cm} Chiral Symmetry,
Chiral Perturbation Theory, Scattering at low energy,
Photon-photon Physics, Meson production.\end{minipage}
\end{center}

\vfill
\begin{abstract}
The processes $\gamma\gamma\to\pi^0\pi^0\pi^0$ and  
$\gamma\gamma\to \pi^+ \pi^- \pi^0$ are considered  in Generalized
Chiral Perturbation theory, in view of their potential sensitivity to the mechanism of spontaneous breaking of chiral symmetry and to various counterterms. 
The amplitudes are computed up to order ${\cal O}({\mbox p}^6)$. 
The event production rates are estimated for the Daphne $\phi$--Factory 
and for a future
$\tau$--Charm Factory.

\end{abstract}
\vfill
\end{titlepage}

\setcounter{footnote}{0}
\renewcommand{\thefootnote}{\arabic{footnote}}

\section{\bf Introduction}

In the limit where the masses of the lightest quark flavours $u,d$ and $s$ are set to zero, the QCD lagrangian becomes invariant under a chiral 
$SU(3)_{\mbox{\scriptsize L}}\times SU(3)_{\mbox{\scriptsize R}}$ global symmetry. This symmetry is not reproduced by the hadronic spectrum, and must therefore be spontaneously broken towards the diagonal $SU(3)_{\rm V}$ subgroup. Actually, this spontaneous breaking of chiral symmetry can be shown to follow from very general properties of QCD \cite{VaWi,tHo,Presk}. However, besides the existence of eight massless pseudoscalar states coupling to the eight conserved axial currents, nothing is known from ``first principles'' about the actual mechanism of spontaneous breaking of chiral symmetry. The widespread belief in that matter is that it proceeds through the formation of a strong quark--antiquark condensate, $<{\bar q}q>\sim 
-(250\,{\mbox MeV})^3$, where $<{\bar q}q>$ denotes the single flavour 
quark--antiquark condensate in the $SU(2)$ chiral limit,
\eq
<{\bar q}q> \,=\, <{\bar u}u>\vert_{m_u=m_d=0}\,=\, <{\bar d}d>\vert_{m_u=m_d=0}\ .
\en
In particular, this means that once the light quark masses $m_u$, $m_d$  and $m_s$ are turned on, the mass of the pion is assumed to be dominated by the contribution linear in quark masses \cite{gor},
\eq
-\frac{2\mhat <{\bar q}q>}{F^2M_{\pi}^2}\sim 1\ ,\ \ \mhat \,=\,\frac{m_u + m_d}{2}\ ,\lbl{stdcond}
\en 
where $F$ stands for the pion decay constant $F_\pi$ (the normalization we use corresponds to the numerical value $F_\pi$ = 92.4 MeV \cite{PDG}) in the same two--flavour chiral limit,
\eq
F = F_{\pi}\vert_{m_u=m_d=0}.
\en
However, our present theoretical knowledge of non--perturbative aspects in QCD does not exclude a picture where the condensate would be much smaller, say $<{\bar q}q>\sim -(100\,{\mbox MeV})^3$, or even vanishing. How the latter possibility may arise in QCD has been discussed recently~\cite{stern98,stern97} in terms of spectral properties of the Dirac operator. On the other hand, it has also
been suggested \cite{KKS98} that a strictly vanishing condensate in the chiral limit could be excluded by an inequality \cite{comellas95} between the correlator $<A_{\mu}(x)A^{\mu}(0)>$ of two axial currents and the two--point function $<P(x)P(0)>$ of the pseudoscalar quark bilinear density $P(x)\equiv ({\bar q}i\gamma_5q)(x)$. This inequality, however, has only been established so far for bare quantities, {\it i.e.} in the presence of an ultraviolet cut--off $\Lambda_{\mbox{\scriptsize UV}}$. As the cut--off is removed, the pseudoscalar density  $P(x)$ needs to be renormalized, whereas the (partially) conserved axial current $A_{\mu}(x)$ remains unaffected. Stricktly speaking, the claim of Ref.~\cite{KKS98} is that $<{\bar q}q>(\Lambda_{\mbox{\scriptsize UV}})\neq 0$ for finite $\Lambda_{\mbox{\scriptsize UV}}$, which does not, {\it a priori}, exclude the possibility of a vanishing condensate in the limit $\Lambda_{\mbox{\scriptsize UV}}\to\infty$ \cite{stern98}. Finally, Ref!
.~\cite{kneur} provides an example of a different approach which, within a variational framework, leads to a non--vanishing, but nevertheless small, value of $<{\bar q}q>$.

Clearly, in order to settle the question of the mechanism of spontaneous chiral symmetry breaking on a purely theoretical level, a major breakthrough in our understanding of non--perturbative aspects of confining gauge theories is required. Instead, one may try a phenomenological approach, and look for experimental observables which could provide the relevant information. In this respect, low--energy $\pi$--$\pi$ scattering in the S--wave has been put forward as a process particularly sensitive to the size of the condensate \cite{pionscat,pipipaper1} (for recent discussions, see {\it e.g.} \cite{pionrev,stern97}). 
This can be most conveniently seen in the framework of the effective lagrangian \cite{weinberg78,GL1,GL2}. In order to incorporate the possibility of a small condensate, the usual counting has however to be modified. As explained in 
Refs.~\cite{pionscat,gchpt}, a consistent expansion scheme, usually refered to as Generalized Chiral Perturbation Theory (GChPT), is obtained by taking (from now on, we restrict ourselves to the case of two massless flavours only),
\eq
m_u,m_d\sim{\cal O}({\mbox p})\ ,\ \ B\sim{\cal O}({\mbox p})\ ,\lbl{gencount}
\en
with p being a generic momentum, much smaller than the typical hadronic scale $\Lambda_H\sim 1\,{\mbox GeV}$, and 
\eq
B\equiv -\frac{<{\bar q}q>}{F^2}\ .
\en
With this counting, the effective lagrangian at lowest order, which consists of all chiral invariant terms of order ${\cal O}({\mbox p}^2)$, reads 
\cite{pionscat}
\bea
{\tilde{\cal L}}^{(2)}&=&{1\over 4}F^2
\left\{\langle D_\mu U^+D^\mu U\rangle +2B\langle U^+\chi+\chi^+U\rangle
\right.\nonumber\\
&&\qquad + A\langle (U^+\chi)^2+(\chi^+U)^2\rangle
 + Z^P\langle U^+\chi-\chi^+U\rangle ^2 \lbl{L2}\\
&&\qquad +\left. h_0\langle \chi^+\chi\rangle 
+ h_1({\mbox{det}}\chi + {\mbox{det}}\chi^+)\right\}\ .\nonumber
\ena
The matrix $U$ collects the pion fields (throughout, we adopt the Condon and 
Shortley phase convention),
\eq
U\,=\, e^{i\phi/F}\ ,\ \ \phi = \left(
\begin{array}{cc}
\pi^0 & -\sqrt{2}\pi^+\\
\sqrt{2}\pi^- & -\pi^0
\end{array}\right)\ .\lbl{Ufield}
\en
The notation is as in Refs.~\cite{GL1,GL2}, except that $\chi$, the quantity 
that contains the scalar and pseudoscalar sources, is defined without the 
usual factor $2B$, 
\eq
\chi={s}+i{ p}={\cal M}+\cdots \ ,\, {\cal M}={\mbox{diag}}(m_u,m_d)
\ .\lbl{chi}
\en
The covariant derivative contains the external vector and axial sources,
\eq
D_{\mu}U\ =\ \partial_{\mu}U-i[v_{\mu},U]-i\{a_{\mu},U\}\ .
\en
This is to be contrasted with Standard Chiral Perturbation Theory (SChPT)~\cite{GL1,GL2}, which assumes a large condensate, and hence the counting rule 
$m_u,m_d\sim{\cal O}({\mbox p}^2)$, $B\sim\Lambda_H$, so that only the first two terms on the right--hand side of Eq.~\rf{L2} are taken into account at leading order. It easily follows from~\rf{L2} that the lowest order expression of the pion mass is now given by \footnote{We neglect the mass difference $m_u -m_d$ and set $m_u=m_d=\mhat$.}
\eq
M_{\pi}^2\,=\,2\mhat B \,+\, 4\mhat^2A \,+\,\cdots\ ,\lbl{Mpilead}
\en
where the ellipsis stands for higher order corrections.
At the same level of approximation in the chiral expansion, the $\pi$--$\pi$ scattering amplitude can be written as~\cite{pionscat}
\eq
A(s\vert t,u)\,=\,\beta\frac{(s-\frac{4}{3}M_{\pi}^2)}{F^2}\,+\,
\alpha\frac{M_{\pi}^2}{3F^2}\,+\,\cdots\ ,\lbl{Amplead}
\en
with
$\beta=1+{\cal O}(\mhat)$, while at this order the parameter $\al$ is directly related to the $<{\bar q}q>$ condensate through
\eq
\alpha = 4 - 3\bigg({{2\mhat B}\over{M_\pi^2}}\bigg)\,+\,{\cal O}(\mhat)\ .
\lbl{alphalead}
\en
The case~\rf{stdcond} of a strong condensate corresponds to $\alpha =1$, whereas the extreme limit where the condensate would vanish yields $\alpha = 4$.
Notice that the above expression for $A(s\vert t,u)$ is not affected
by the additional terms in \rf{L2} and reproduces the result first obtained by Weinberg~\cite{wein66}, $A(s\vert t,u)\,=\,(s-2\mhat B)/F^2\,+\,\cdots$. The difference between the standard case and deviations from it lies here only in the leading--order expression~\rf{Mpilead} of the pion mass and its relation to the condensate. Higher orders in the chiral expansion will modify the simple expression \rf{Amplead}, but the correlation between low--energy $\pi$--$\pi$ scattering and the value of the ratio $2\mhat B/M_{\pi}^2$ subsists, and can be studied in a controled way within the generalized chiral expansion \cite{pipipaper1}. Available data on low--energy $\pi$--$\pi$ phases, which are dominated by the Geneva--Saclay $K_{\ell 4}$ experiment~\cite{ross77}, do however not possess the required accuracy in order to distinguish between the different alternatives at present. Forthcoming experiments, such as new $K_{\ell 4}$ experiments, conducted by the BNL865 collaboration \cite{b!
nl865} or planed at the Daphne $\phi$--factory \cite{daphne}, and the DIRAC experiment at CERN 
\cite{dirac}, represent promissing prospects in this direction.

In order to create a (hopefully convergent) set of evidences {\it pro} or {\it contra} a specific picture of spontaneous breaking of chiral symmetry, it remains however important to explore other possibilities, and to find other processes which, at the theoretical level, can be shown to exhibit a reasonably strong dependence on the value of the condensate. The present work was motivated by the above consideration and the following observation. A straightforward re--analysis in GChPT of the existing SChPT calculations~\cite{adler,bos,pere96} at lowest order shows that the two amplitudes for the production of three pions in low--energy photon--photon collisions  involve the parameter $\alpha$ already at tree level.
In the case of $\gamma\gamma\to\pi^0\pi^0\pi^0$, the amplitude reads 
\eq
{\cal A}^{N} =
{{e^2}\over{4\pi^2F_\pi^3}}\epsilon^{\mu\nu\al\be}k_\mu\epsilon_\nu k_\al'\epsilon_\be'
\,{{\al M_\pi^2}\over{s-M_\pi^2}}\,+\,\cdots\ ,
\en
while for $\gamma\gamma\to \pi^+\pi^-\pi^0$ we find
\bea
{\cal A}^{C} &=&
{{e^2}\over{4\pi^2F_\pi^3}}\epsilon_{\mu\nu\al\be}k^\mu\epsilon^\nu k'^\al\epsilon'^\be\,\bigg(1-{{(p_++p_-)^2-M_\pi^2+{1\over 3}(\al -1)M_\pi^2}\over
{s-M_\pi^2}}\bigg)
\nonumber\\
&&
+{{e^2}\over{4\pi^2F_\pi^3}}
\epsilon_{\mu \nu \alpha \beta} k'^\mu \epsilon'^\nu
p_0^{\alpha} \Bigl(-\epsilon^\beta+ {\epsilon \cdot p_{-}\over
k\cdot p_{-}}p_+^{\beta} +  {\epsilon \cdot p_{+}\over
k\cdot p_{+}}p_-^{\beta}\Bigr)
\nonumber\\
&&
+{{e^2}\over{4\pi^2F_\pi^3}}
\epsilon_{\mu \nu \alpha \beta} k^\mu \epsilon^\nu
p_0^{\alpha} \Bigl(-\epsilon'^\beta + {\epsilon' \cdot p_{-}\over
k'\cdot p_{-}}p_+^{\beta} 
+  {\epsilon' \cdot p_{+}\over 
k'\cdot p_{+}}
p_-^\beta\Bigr)\,+\,\cdots\ .
\ena
In these expressions, $k$ and $k'$ ( $\epsilon$ and $\epsilon'$) denote the momenta (polarizations) of the two photons, while $p_+$, $p_-$ and $p_0$ are the pion momenta. The ellipses stand for higher order corrections which will be considered later. Diagrammatically, the origin of the dependence on $\alpha$ lies in the presence of one--pion reducible contributions to the two amplitudes, see Fig.~1 below.

\indent

\centerline{\psfig{figure=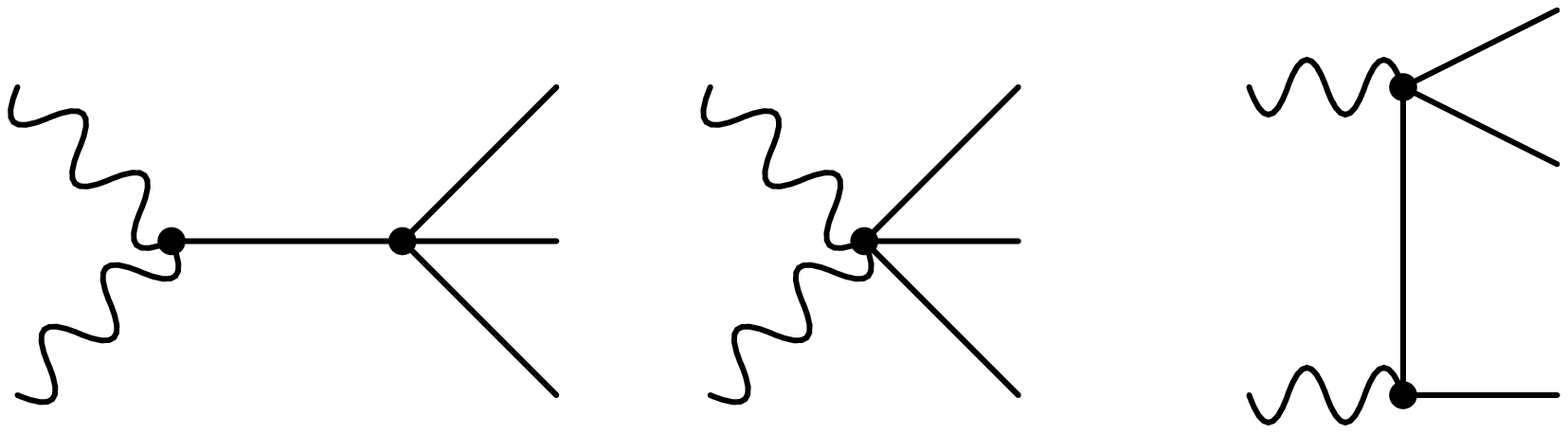,height=3.5cm}}

\noindent
{\bf Figure 1} : {\it The lowest order contributions to the $\gamma\gamma\to\pi\pi\pi$ amplitudes. The $\alpha$--dependence in ${\cal A}^{N}$ and  ${\cal A}^{C}$ comes from the vertex for (virtual) $\pi$--$\pi$ scattering in the first graph, which is the only one to contribute in the case of the neutral amplitude.}

\indent

\noindent
Therefore, the leading--order neutral amplitude ${\cal A}^N$ increases as the condensate decreases. For a strictly vanishing condensate ($\al =4$), the cross section at low energies is thus enhanced by a factor 16 as compared to the standard case of a strong condensate ($\al =1$) ! In the charged case, the situation is less favourable. At threshold, the average over the photon polarizations of the modulus squared of the amplitude is still proportional to the square of 
$\al$,
\eq
{1\over 4}\sum_{\mbox{\scriptsize pol}}\vert{\cal A}^C
\vert^2_{\mbox{\scriptsize thr}}
={1\over 4}\,\bigg({{e^2}\over{4\pi^2F_\pi^3}}\bigg)^2\,
{{9M_\pi^4}\over 128}\,\al^2\ ,
\en
but the sensitivity on $\al$ of the corresponding total {\it cross section} $\sigma^{C}(s,\al)$ rapidly decreases with increasing energy. For instance, the ratio $\sigma^C(s,\al=3)/\sigma^C(s,\al=1)$ is equal to 4.48 at $\sqrt{s}=450$ MeV, {\it i.e.} just above threshold, but drops to 1.68 at $\sqrt{s}=500$ MeV and becomes less than 1.10 at $\sqrt{s}\gapprox 600$ MeV. For the case $\al =2$, 
the corresponding ratio is equal to 2.21 at $\sqrt{s}=450$ MeV, but the effect is less 
than 25\% at $\sqrt{s}=500$ MeV, while it barely reaches a few percent at 
$\sqrt{s}\gapprox 600$. Thus, the interference between the two kinematical structures contributing to the amplitude
${\cal A}^C$, which is responsible, at low energies, for the suppression of the cross section in the charged channel as compared to the cross section in the neutral channel~\cite{pere96}, also washes out the dependence on the value of the $<{\bar q}q>$ condensate as soon as one leaves the threshold region. Considering definite polarization configurations for the two photons does not improve the situation~: For parallel polarizations of the two photons, the part of ${\cal A}^C$ which is sensitive to $\alpha$ does not contribute to the cross section, whereas for polarizations taken along orthogonal axes, the same interference effect is again fully at work. Furthermore, both cross sections rapidly rise above threshold, so that the effect of higher orders also needs to be investigated. The purpose of this paper is precisely to investigate these higher order effects at the one--loop level in GChPT. In fact, at next--to--leading order new tensorial structures appear both in the neut!
ral and in the charged amplitude \cite{pere96}, whereas the pion loops provide additional sources of dependence with respect to $\alpha$, so that their behaviour with respect to changes in the value of the condensate could be modified to some extent. 

Accordingly, the rest of the paper is organized as follows~: The general kinematical structure of the two amplitudes ${\cal A}^N$ and ${\cal A}^C$, as well as their properties under isospin symmetry, are the subject of Section 2. The construction and the renormalization of the effective lagrangian of GChPT at 
order one--loop in the two--flavour case are treated in Section 3. Section 4 is devoted to the actual calculation of the two amplitudes to one--loop precision. The counterterms which are involved in these expressions and several numerical results are discussed in Section 5. A summary and conclusions are presented in Section 6. Details on the evaluation of some counterterms have been gathered in 
an Appendix.

\indent
\setcounter{equation}{0}

\section{Kinematics and Isospin Symmetry}
\setcounter{equation}{0}

The amplitudes ${\cal A}^N$ and ${\cal A}^C$ for the processes

\bea
\gamma(k)\ \gamma(k')\ &\to & \pi^0(p_1)\ \pi^0(p_2)\ \pi^0(p_3)\ ,
\nonumber\\
\gamma(k)\ \gamma(k')\ &\to & \pi^+(p_+)\ \pi^-(p_-)\ \pi^0(p_0)\ ,\lbl{proc}
\ena
are obtained from the matrix elements 
\bea
<\,\pi^0(p_1)\pi^0(p_2)\pi^0(p_3)\,{\mbox{out}}\,\vert\,
\gamma(k,\epsilon)\gamma(k',\epsilon')\,{\mbox{in}}\,>&=&
i(2\pi)^4\delta^4(P_f-P_i){\cal A}^N\ ,\nonumber\\
<\,\pi^+(p_+)\pi^-(p_-)\pi^0(p_0)\,{\mbox{out}}\,\vert\,
\gamma(k,\epsilon)\gamma(k',\epsilon')\,{\mbox{in}}\,>&=&
i(2\pi)^4\delta^4(P_f-P_i){\cal A}^C\ , \lbl{mat} 
\ena
respectively, with
\bea
&&
{\cal A}^N(k,\epsilon;k',\epsilon';p_1,p_2,p_3)
\,=\,
\nonumber\\
&&
\qquad
ie^2\epsilon_{\mu}(k)\epsilon_{\nu}'(k')
\int d^4x e^{-ik\cdot x}
<\,\pi^0(p_1)\pi^0(p_2)\pi^0(p_3)\,{\mbox out}\,\vert\,
T\{j^{\mu}(x)j^{\nu}(0)\}\,\vert\,\Omega\,>\ ,
\nonumber\\
&&
{\cal A}^C(k,\epsilon;k',\epsilon';p_0,p_+,p_-)
\,=\,
\nonumber\\
&&
\qquad
ie^2\epsilon_{\mu}(k)\epsilon_{\nu}'(k')
\int d^4x e^{-ik\cdot x}
<\,\pi^+(p_+)\pi^-(p_-)\pi^0(p_0)\,{\mbox out}\,\vert\,
T\{j^{\mu}(x)j^{\nu}(0)\}\,\vert\,\Omega\,>\ .\lbl{red}
\ena
In the above expressions, $j_{\mu}(x)$ denotes the electromagnetic
current, with its usual decomposition into an isotriplet and an isosinglet component, $j_{\mu}=j_{\mu}^3+j_{\mu}^0$, while $\vert\,\Omega\,>$ stands for the QCD vacuum with massive light quarks, but in the absence of electromagnetism (radiative corrections to the two processes \rf{proc} are not considered here). Since we also neglect isospin breaking effects due to $m_u\neq m_d$, Bose symmetry and G--parity constrain the three final pions to be in an $I=1$ total isospin state. Thus the matrix elements in
Eq. \rf{red} only involve the $I=1$ component of the product of the two 
electromagnetic currents,
\eq
T\big\{j_{\mu}(x)j_{\nu}(0)\big\}_{I=1}\ =
\ T\big\{j^3_{\mu}(x)j^0_{\nu}(0)+j^0_{\mu}(x)j^3_{\nu}(0)\big\}\ . 
\en
Therefore, isospin invariance relates the two amplitudes ${\cal A}^C$ and ${\cal A}^N$ in a simple way. With the Condon and Shortley phase convention adopted in Eq.~\rf{Ufield}, this relation reads
\eq
-{\cal A}^N(k,\epsilon;k',\epsilon';p_1,p_2,p_3)\ =\ 
{\cal A}^C(k,\epsilon;k',\epsilon';p_1,p_2,p_3)+ {\mbox{cyclic}}\,(p_1,p_2,p_3)\ ,
\lbl{iso}
\en
where ``${\mbox{cyclic}}\,(p_1,p_2,p_3)$'' indicates that the contributions arising from cyclic permutations over the pion momenta $p_1, p_2$ and $p_3$  have to be added.

Up to permutations of the momenta and polarizations of the photons, and/or permutations of the momenta of the charged pions, the amplitude ${\cal A}^C$  may be decomposed into six independent Lorentz invariant amplitudes
\bea
&&{\cal A}^C(k,\epsilon;k',\epsilon';p_0,p_+,p_-) \ =
\ {\cal A}^C_1(k,k';p_0,p_+,p_-)\,
t_1(k,\epsilon;k',\epsilon')\nonumber\\
&&\quad
+\ \sum_{i=2,3}\,\bigg[{\cal A}^C_i(k,k';p_0,p_+,p_-)\,
t_i(k,\epsilon;k',\epsilon';p_0,p_+,p_-)\ +
\ { k \choose \epsilon } \leftrightarrow 
{ k' \choose \epsilon' } \bigg]\lbl{decomp}\\
&&\quad
+\ \sum_{i=4,5,6}\,\bigg\{\bigg[{\cal A}^C_i(k,k';p_0,p_+,p_-)\,
t_i(k,\epsilon;k',\epsilon';p_0,p_+,p_-)\ +
\  { k \choose \epsilon } \leftrightarrow 
{ k' \choose \epsilon' } \bigg] +
\bigg[p_{+} \leftrightarrow p_{-} \bigg]\bigg\}
\ ,\nonumber
\ena
with ($p_{ij}\equiv p_i+p_j$, $i,j=+,-,0$)
\bea
& & t_1(k,\epsilon;k',\epsilon')=
\epsilon_{\mu \nu \alpha \beta} k^\mu \epsilon^\nu
k'^\alpha \epsilon'^\beta, \nonumber \\
& & t_2(k,\epsilon;k',\epsilon';p_0,p_+,p_-)=\epsilon_{\mu \nu \alpha \beta} k'^\mu \epsilon'^\nu
p_{0}^\alpha \Bigl(-\epsilon^\beta+ {\epsilon \cdot p_{-}\over
k\cdot p_{-}}p_{+}^{\beta} +  {\epsilon \cdot p_{+}\over
k\cdot p_{+}}p_{-}^{\beta}\Bigr), \nonumber \\
& & t_3(k,\epsilon;k',\epsilon';p_0,p_+,p_-)=\epsilon_{\mu \nu \alpha \beta}
\Bigl(-\epsilon^\mu +{\epsilon \cdot p_{+-} \over k\cdot p_{+-}}k^\mu\Bigr)
k'^\nu p_{0}^\alpha
\epsilon'^\beta,
 \nonumber \\
& & t_4(k,\epsilon;k',\epsilon';p_0,p_+,p_-)=\epsilon_{\mu \nu \alpha \beta}
\Bigl(-\epsilon^\mu +{\epsilon \cdot p_{+0} \over k\cdot p_{+0}}k^\mu\Bigr)
k'^\nu p_{-}^\alpha
\epsilon'^\beta,
 \nonumber \\
& & t_5(k,\epsilon;k',\epsilon';p_0,p_+,p_-)=\epsilon_{\mu \nu \alpha \beta}
k'^\mu p_{+}^\nu p_{-}^\alpha \epsilon'^\beta \Bigl({\epsilon \cdot p_{+0}
\over k \cdot p_{+0}} - {\epsilon \cdot p_{+} \over k \cdot p_{+}}\Bigr),
 \nonumber \\
& &t_6(k,\epsilon;k',\epsilon';p_0,p_+,p_-) =\epsilon_{\mu \nu \alpha \beta}
\Bigl(-\epsilon^\mu +{\epsilon \cdot p_{+} \over k\cdot
p_{+}}k^\mu \Bigr) k'^\nu p_{-}^\alpha \epsilon'^\beta\ .
\ena

In the sequel, we shall compute the amplitudes ${\cal A}^C$ and ${\cal A}^N$ within the framework of generalized chiral perturbation theory up to order one--loop. At this level of accuracy of the chiral expansion, the amplitude ${\cal A}^N$ does not yet receive its full structure as implied by Eqs. \rf{decomp} and \rf{iso}. Rather, it takes the simpler form ($p_{ij}\equiv p_i+p_j$, $i,j=1,2,3$)
\bea
&&{\cal A}^N(k,\epsilon;k'\epsilon';p_1,p_2,p_3) \ =
\ {\cal A}^N_1(k,k';p_1,p_2,p_3)\,
t_1(k,\epsilon;k',\epsilon')
\nonumber\\
&&\quad
+\ \bigg[{\cal A}^N_2(k,k';p_1,p_2,p_3)
\epsilon_{\mu \nu \alpha \beta}
\Bigl(\epsilon'^\mu -{\epsilon' \cdot p_{12} \over k'\cdot p_{12}}k'^\mu\Bigr)
p_{12}^{\nu}\epsilon^{\alpha}k^{\beta}\,+\,{\mbox{cyclic}}\,(p_1,p_2,p_3)\bigg]
\nonumber\\
&&\quad +
\ \bigg[{ k \choose \epsilon } \leftrightarrow 
{ k' \choose \epsilon' } \bigg]\ +\ \cdots\ ,\lbl{Aneu}
\ena
where the ellipsis stands for higher order terms in the chiral expansion. In order that \rf{Aneu} follows from \rf{iso} and \rf{decomp}, it is {\it sufficient} that the one--loop charged amplitudes ${\cal A}_i^C$ satisfy the following conditions~:
\begin{description}
\item[i)]
${\cal A}_2^C(k,k';p_0,p_+,p_-)-{\cal A}_5^C(k,k';p_+,p_-,p_0)$ is entirely symmetric under permutations of the pion momenta $p_+, p_-, p_0$;
\item[ii)]
$k\cdot p_0\,\big[{\cal A}_5^C(k,k';p_0,p_+,p_-)+
{\cal A}_6^C(k,k';p_0,p_+,p_-)\big]\ =$

$\qquad\ k\cdot p_+
\,\big[{\cal A}_5^C(k,k';p_+,p_0,p_-)+
{\cal A}_6^C(k,k';p_+,p_0,p_-)\big]$;
\item[iii)]
${\cal A}_5^C(k,k';p_0,p_+,p_-)\ =\ {\cal A}_5^C(k,k';p_+,p_0,p_-)$.
\end{description}
Furthermore, the two neutral amplitudes have then the following expressions in 
terms of the charged ones~:
\bea
{\cal A}^N_1(k,k';p_1,p_2,p_3) &=&
-{\cal A}^C_1(k,k';p_1,p_2,p_3)\nonumber\\
&&+\ \bigg[\,{1\over 3}{\cal A}^C_2(k,k';p_1,p_2,p_3)+
{\cal A}^C_3(k,k';p_3,p_1,p_2)\nonumber\\
&&+{\cal A}^C_4(k,k';p_1,p_2,p_3)+
{\cal A}^C_4(k,k';p_2,p_1,p_3)\nonumber\\
&&+{\cal A}^C_6(k,k';p_1,p_2,p_3)+
{\cal A}^C_6(k,k';p_2,p_1,p_3)\nonumber\\
&&+{2\over 3}{\cal A}^C_5(k,k';p_2,p_3,p_1)\,\bigg]\ +
\ \bigg[{ k \choose \epsilon } \leftrightarrow 
{ k' \choose \epsilon'} \bigg]\nonumber\\
&&+\ {\mbox{cyclic}}\,(p_1,p_2,p_3)\ ,\lbl{iso1}
\ena
and
\bea
-{\cal A}^N_2(k,k';p_1,p_2,p_3) &=&
\big[
{\cal A}^C_4(k,k';p_1,p_2,p_3)+{\cal A}^C_6(k,k';p_1,p_2,p_3)\big]+
\big[\,p_1\leftrightarrow p_2\,\big]\nonumber\\
&&+
{\cal A}^C_3(k,k';p_1,p_2,p_3)+{\cal A}^C_5(k,k';p_2,p_1,p_3)\ .\lbl{iso2}
\ena

\section{The Effective Lagrangian in GChPT}
\setcounter{equation}{0}

Before proceeding with the calculation of the amplitudes ${\cal A}^C$ and ${\cal A}^N$ in the next section, we first discuss the structure of the low--energy generating functional in the case of two light flavours that we need for our subsequent calculation. Since most of the results of this section are not available from the existing literature, we discuss them in some detail.

The structure of the effective lagrangian ${\cal L}^{\mbox{\scriptsize eff}}$ is independent of the underlying mechanism of spontaneous chiral symmetry breaking. It consists of an infinite tower of chiral invariant contributions

\eq
{\cal L}^{\mbox{\scriptsize eff}} = \sum_{(k,l)}\,{\cal L}_{(k,l)}\ ,\lbl{Leff}
\en
where ${\cal L}_{(k,l)}$ contains $k$ powers of covariant derivatives and $l$ powers of the scalar or pseudoscalar sources. In the chiral limit, these terms behave as
\eq
{\cal L}_{(k,l)}\sim\left({{\mbox p}\over{\Lambda_H}}\right)^k
\left({{m_{\mbox{\scriptsize quark}}}\over{\Lambda_H}}\right)^l\ ,
\en
with $m_{\mbox{\scriptsize quark}}=m_u,m_d$, and p stands for a typical external momentum.
The standard approach not only assumes 
$m_{\mbox{\scriptsize quark}}\ll\Lambda_H$, but also 
$m_{\mbox{\scriptsize quark}}\ll B/2A$, 
such as to enforce the dominance of the term linear in 
$m_{\mbox{\scriptsize quark}}$ in the 
expression of the pion mass~\rf{Mpilead}. This allows to reorganize the double
 expansion \rf{Leff} as~\cite{GL1,GL2}
\eq
{\cal L}^{\mbox{\scriptsize eff}} =  
{\cal L}^{(2)}+{\cal L}^{(4)}+{\cal L}^{(6)}+\cdots\ ,
\en
where ${\cal L}^{(d)}=\sum\,{\cal L}_{(k,l)}$ with $k+2l=d$.
The generalized framework considers the possibility that the condensate could be much smaller than usually believed, so that for the actual values of the quark masses one could have $m_{\mbox{\scriptsize quark}}\sim B/2A$ and still 
$m_{\mbox{\scriptsize quark}}\ll\Lambda_H$. This leads to a different 
reorganization of the double expansion \rf{Leff}, namely
\eq
{\cal L}^{\mbox{\scriptsize eff}} =  
{\tilde{\cal L}}^{(2)}+{\tilde{\cal L}}^{(3)}+{\tilde{\cal L}}^{(4)}
+{\tilde{\cal L}}^{(5)}+{\tilde{\cal L}}^{(6)}+\cdots\ ,
\en
where now, according to \rf{gencount}, ${\tilde{\cal L}}^{(d)}=\sum\,B^n{\cal L}_{(k,l)}$ with $k+l+n=d$ \cite{pionscat,gchpt}.

The leading order of the generalized expansion is described by ${\tilde{\cal L}}^{(2)}$, which in the two flavour case was given in Section 1, Eq.~\rf{L2}. For our purposes, we  need only to consider the situation without axial source, and with the vector source restricted to the (classical) photon field $A_{\mu}$,
\eq
a_{\mu}\ =\ 0\ ,\ \ v_{\mu}\ =\ eA_{\mu}Q\ ,
\en
where $Q$ stands for the charge matrix of the two light quark flavours $u$ and $d$,
\eq
Q\ =\ \left( \begin{array}{cc}
{2\over 3} & 0\\
0 & -{1\over 3}\end{array}\right)\ .
\en

In GChPT, the next--to--leading--order corrections are of order 
${\cal O}({\mbox p}^3)$, and still occur before the loop corrections. They are
embodied in ${\tilde{\cal L}}^{(3)}={\cal L}_{(2,1)}+{\cal L}_{(0,3)}$, 
which reads~\footnote{~The superscript (2) is meant to distinguish the low energy constants $\xi^{(2)}$, $\rho_i^{(2)}$ from the similar ones that occur in the expression of ${\tilde{\cal L}}^{(3)}$ in the three flavour case 
\cite{pionscat,gchpt}.} 
\bea
\tilde{\cal L}^{(3)}&=&{1\over 4}F^2
\left\{\xi^{(2)}\langle D_\mu U^+D^\mu U(\chi^+U+U^+\chi)\rangle 
\right.\nonumber\\
&&\qquad + \rho_1^{(2)}\langle (\chi^+U)^3+(U^+\chi)^3\rangle
+\rho_2^{(2)}\langle (\chi^+U+U^+\chi)\chi^+\chi\rangle\nonumber\\
&&\qquad + \rho_3^{(2)}\langle\chi^+U-U^+\chi\rangle
\langle(\chi^+U)^2-(U^+\chi)^2\rangle\lbl{L3} \\
&&\qquad + \rho_4^{(2)}\langle(\chi^+U)^2
+(U^+\chi)^2\rangle
\langle\chi^+U+U^+\chi\rangle\nonumber\\
&&\qquad \left. + \rho_5^{(2)}\langle\chi^+\chi\rangle
\langle\chi^+U+U^+\chi\rangle
\right\}\ .\nonumber
\ena

>From order ${\cal O}({\mbox p}^4)$ onward, the contributions to 
${\cal L}^{\mbox{\scriptsize eff}}$ come with either even or odd intrinsic 
parity, ${\tilde{\cal L}}^{(d)} = {\tilde{\cal L}}^{(d)}_+
+{\tilde{\cal L}}^{(d)}_-$ for $d\ge 4$, or 
${\cal L}_{(k,l)}={\cal L}_{(k,l)}^{+}+{\cal L}_{(k,l)}^{-}$ for $k\ge 4$.
The tree--level contributions at order ${\cal O}({\mbox p}^4)$ in the {\it even intrinsic parity sector} are contained in 
\eq
\tilde{\cal L}^{(4)}_+={\cal L}_{(4,0)}^{+}+{\cal L}_{(2,2)}+{\cal L}_{(0,4)}+
B^2{\cal L}'_{(0,2)} + B{\cal L}'_{(2,1)} + B{\cal L}'_{(0,3)}\ .
\lbl{L4}
\en
The part without explicit chiral symmetry breaking, ${\cal L}_{(4,0)}^{+}$, 
is described by the same low--energy constants
$l_1$, $l_2$, $l_5$, $l_6$ and $h_2$ as in the standard case  \cite{GL1},
\bea
{\cal L}_{(4,0)}^{+} &=& {{l_1}\over 4}\,\langle\,D^{\mu}U^+D_{\mu}U\,
\rangle^2\,+\,{{l_2}\over 4}\,\langle\,D^{\mu}U^+D^{\nu}U\,
\rangle\,\langle\,D_{\mu}U^+D_{\nu}U\,\rangle
\nonumber\\
&&
+\,l_5\,\langle
\,F^{\mbox{\scriptsize L}}_{\mu\nu}UF^{{\mbox{\scriptsize L}}\,\mu\nu}U^+
\,\rangle
\,+\,{{il_6}\over 2}\,\langle
\,F^{\mbox{\scriptsize R}}_{\mu\nu}d^{\mu}Ud^{\nu}U^+ 
+ F^{\mbox{\scriptsize L}}_{\mu\nu}d^{\mu}U^+d^{\nu}U\,\rangle
\nonumber\\
&&
-\,(2h_2+\frac{1}{2}l_5)\,\langle
\,F^{\mbox{\scriptsize R}}_{\mu\nu}F^{{\mbox{\scriptsize R}}\,\mu\nu}
+F^{\mbox{\scriptsize L}}_{\mu\nu}F^{{\mbox{\scriptsize L}}\,\mu\nu}
\,\rangle\ .
\ena
 The part with two powers of momenta and two powers of quark masses is given 
by
\bea
{\cal L}_{(2,2)}&=& {1\over{4}}F^2\bigg\{
a_1 \langle D_{\mu}U^+D^{\mu}U (\chi^+\chi + U^+\chi\chi^+U)\rangle 
                                                            \nonumber\\    
& &\qquad + a_2 \langle D_{\mu}U^+U\chi^+D^{\mu}UU^+\chi\rangle \nonumber\\
& &\qquad + a_3 \langle D_{\mu}U^+U(\chi^+D^{\mu}\chi-D^{\mu}\chi^+\chi) +
          D_{\mu}UU^+(\chi D^{\mu}\chi^+ - D^{\mu}\chi\chi^+)\rangle 
\nonumber\\
& &\qquad + b_1 \langle D_{\mu}U^+D^{\mu}U (\chi^+U\chi^+U + 
                                        U^+\chi U^+\chi)\rangle \nonumber\\
& &\qquad + b_2 \langle D_{\mu}U^+\chi D^{\mu}U^+\chi + 
                        \chi^+D_{\mu}U\chi^+D^{\mu}U\rangle\nonumber\\ 
& &\qquad + b_3 \langle
U^+D_{\mu}\chi U^+D^{\mu}\chi+D_{\mu}\chi^+UD^{\mu}\chi^+U\rangle\nonumber\\
& &\qquad + c_1 \langle D_{\mu}U^+\chi +\chi^+ D_{\mu}U\rangle
                \langle D^{\mu}U^+\chi +\chi^+ D^{\mu}U\rangle\nonumber\\
& &\qquad + c_2 \langle D_{\mu}\chi^+U + U^+ D_{\mu}\chi\rangle
                \langle D^{\mu}U^+\chi +\chi^+ D^{\mu}U\rangle\nonumber\\
& &\qquad + c_3 \langle D_{\mu}\chi^+U + U^+ D_{\mu}\chi\rangle
                \langle D^{\mu}\chi^+U + U^+D^{\mu}\chi\rangle \lbl{L22}\\
& &\qquad + c_4 \langle D_{\mu}U^+\chi -\chi^+ D_{\mu}U\rangle
                \langle D^{\mu}U^+\chi -\chi^+ D^{\mu}U\rangle\nonumber\\
& &\qquad + c_5 \langle D_{\mu}\chi^+U - U^+D_{\mu}\chi\rangle
                \langle D^{\mu}\chi^+U - U^+D^{\mu}\chi\rangle\nonumber\\
& &\qquad + h_3 \langle D_{\mu}\chi^+D^{\mu}\chi^+\rangle 
                               \bigg\} \ .\nonumber
\ena
Finally, the tree--level contributions which behave as ${\cal O}
(m_{\mbox{\scriptsize quark}}^4)$ in 
the chiral limit are contained in ${\cal L}_{(0,4)}$, which reads
\bea
{\cal L}_{(0,4)} &=& {1\over{4}}F^2\bigg\{
  e_1 \langle (\chi^+U)^4 + (U^+\chi)^4 \rangle \nonumber\\
& &\qquad 
+ e_2 \langle \chi^+\chi(\chi^+U\chi^+U+U^+\chi U^+\chi) \rangle\nonumber\\
& &\qquad 
+ e_3 \langle \chi^+\chi U^+\chi\chi^+ U \rangle \nonumber\\
& &\qquad 
+ f_1 \langle (\chi^+U)^2 + (U^+\chi)^2 \rangle^2\nonumber\\
& &\qquad 
+ f_2 \langle (\chi^+U)^3 + (U^+\chi)^3 \rangle\langle \chi^+U + U^+\chi\rangle
\nonumber\\
& &\qquad  
+ f_3 \langle \chi^+\chi(\chi^+U + U^+\chi)\rangle
                           \langle\chi^+U + U^+\chi\rangle\nonumber\\
& &\qquad 
+ f_4 \langle (\chi^+U)^2 + (U^+\chi)^2 \rangle
      \langle \chi^+U + U^+\chi\rangle^2
\nonumber\\
& &\qquad +
f_5 \langle (\chi^+U)^3 - (U^+\chi)^3 \rangle\langle \chi^+U - U^+\chi\rangle
\nonumber\\
& &\qquad +
h_4 \langle\chi^+\chi\chi^+\chi \rangle\nonumber\\
& &\qquad +
h_5 \langle\chi^+\chi \rangle ({\mbox{det}}\chi + 
{\mbox{det}}\chi^+ )\nonumber\\
& &\qquad +
h_6 ({\mbox{det}}\chi + {\mbox{det}}\chi^+ )^2 \nonumber\\
& &\qquad +
h_7 ({\mbox{det}}\chi - {\mbox{det}}\chi^+ )^2 
 \bigg\} \ . \nonumber
\ena
Notice  that in the
standard framework  the contributions from ${\cal L}_{(2,2)}$ and from ${\cal L}_{(0,4)}$would count as order ${\cal O}({\mbox p}^6)$ and
order ${\cal O}({\mbox p}^8)$, respectively~\footnote{~The standard 
${\cal O}({\mbox p}^6)$ effective lagrangian ${\cal
L}^{(6)}={\cal L}_{(6,0)}+{\cal L}_{(4,1)}+{\cal L}_{(2,2)}+{\cal L}_{(0,3)}$ 
has been worked out in Ref.  \cite{FS} for the case of three light flavours, 
and very recently, for both two and three light flavours, in 
Ref.~\cite{bijnensetal}.}.

Next, we turn to the {\it odd intrinsic parity sector}. There, the first contribution starts at order ${\cal O}({\mbox p}^4)$, and since it is entirely fixed by the short distance properties of QCD, there is no difference between the standard and the generalized case, ${\cal L}^{(4)}_-={\tilde{\cal L}}^{(4)}_-=
{\cal L}_{(4,0)}^-$. In the two flavour case, ${\cal L}_{(4,0)}^-$ vanishes in the absence of external sources. In the presence of an electromagnetic field, it reads
\bea
{\cal L}_{(4,0)}^-&=&
{e\over{16\pi^2}}\epsilon^{\mu\nu\alpha\beta}A_{\mu}\langle
Q\big(\partial_\nu U\partial_\alpha U^+\partial_\beta UU^+-
      \partial_\nu U^+\partial_\alpha U\partial_\beta U^+U\big)\rangle
\\
&&
-{{ie^2}\over{8\pi^2}}\epsilon^{\mu\nu\alpha\beta}\partial_\mu A_\nu A_\alpha
\langle Q^2\partial_\beta UU^++Q^2U^+\partial_\beta U
-{1\over 2}QUQ\partial_\beta U^+ +{1\over 2}QU^+Q\partial_\beta U\rangle\ .
\nonumber
\ena
The computation of the amplitudes ${\cal A}^N$ and ${\cal A}^C$ to one loop also involves the counterterms from ${\tilde{\cal L}}^{(5)}_-={\cal L}_{(4,1)}^-$ and from ${\tilde{\cal L}}^{(6)}_-={\cal L}_{(6,0)}^-+{\cal L}_{(4,2)}^-$. In the standard case, both ${\cal L}_{(4,1)}^-$ and ${\cal L}_{(6,0)}^-$ count as order ${\cal O}({\mbox p}^6)$, and have been discussed before in the literature \footnote{For a review, see \cite{bijnensrev}.} 
\cite{Issl,AkAl,FS}. Borrowing from the last and most recent of these references, we obtain

\bea
{\cal L}_{(4,1)}^- &=& {1\over{4\pi^2}}\epsilon_{\mu\nu\al\be}\bigg\{
iA_{4} \langle\cm{\chi}\cp{G^{\mu\nu}}\cp{G^{\al\be}}\rangle 
+ iA_{6} \langle\cm{\chi}\rangle\langle\cp{G^{\mu\nu}}\cp{G^{\al\be}}\rangle
\nonumber\\
&&
+ A_{12} \langle\cm{D^\mu U}\cm{D^\nu U}(\cm{\chi}\cp{G^{\al\be}} +
\cp{G^{\al\be}}\cm{\chi})\rangle 
\nonumber\\
&&
+ A_{13} \langle\cm{D^\mu U}\cm{\chi}\cm{D^\nu U}\cp{G^{\al\be}}\rangle
\,\cdots\bigg\}\ ,\lbl{L41-}
\ena

and

\bea
{\cal L}_{(6,0)}^- &=& {1\over{4\pi^2}}\epsilon_{\mu\nu\al\be}\bigg\{
 iA_{2} \langle\cm{D^\mu U}(\cp{D^\nu G^{\ga\al}}\cp{{G_\ga}^\be} -
\cp{G^{\ga\al}}\cp{D^\nu {G_\ga}^\be})\rangle
\nonumber\\
&&
+ iA_{3} \langle\cm{D^\mu U}(\cp{D_\ga G^{\ga\nu}}\cp{G^{\al\be}} -
\cp{G^{\ga\nu}}\cp{D_\ga G^{\al\be}} 
\nonumber\\
&&\qquad\qquad\qquad\qquad
-\cp{D_\ga G^{\al\be}}\cp{G^{\ga\nu}} + \cp{G^{\al\be}}\cp{D_\ga G^{\ga\nu}})
\rangle 
\nonumber\\
&&
+ A_{7} \langle\cm{D^\al D^\ga U}(\cm{D_\ga U}\cm{D^\be U}\cp{G^{\mu\nu}} 
-\cp{G^{\mu\nu}}\cm{D^\be U}\cm{D_\ga U})\rangle
\nonumber\\
&&
+ A_{8} \langle\cm{D^\al D^\ga U}(\cm{D^\be U}\cm{D_\ga U}\cp{G^{\mu\nu}} -
\cp{G^{\mu\nu}}\cm{D_\ga U}\cm{D^\be U})\rangle
\nonumber\\
&&
+\ \cdots\bigg\}\ .
\lbl{L60-}
\ena

\noindent
Here, we have only listed those terms that will actually contribute to 
the processes under study, when the mass--shell conditions for the momenta and polarizations of the photons        
are taken into account. We have however kept the numbering of the low-energy constants introduced in \cite{FS}, but we have, for convenience, changed their normalization by an overall factor $1/4\pi^2$. The notation is otherwise as in \cite{FS}, except for the fact that the source $\chi$ does not contain the factor $2B$, see Eq. \rf{chi}. 

The last piece we need for a full one--loop computation of the amplitudes \rf{red} is ${\cal L}_{(4,2)}^-$. It counts as order ${\cal O}({\mbox p}^8)$ in the {\it standard} case, and is not available from the existing literature. These contributions, which are order ${\cal O}(\mhat^2)$ corrections to 
${\cal L}_{(4,0)}^-$, are expected to be small in the two--flavour chiral expansion, and will be parametrized appropriately in the one--loop expressions of the amplitudes ${\cal A}^C$ and ${\cal A}^N$ given in the next section. The determination of the combinations of low--energy constants that enter these amplitudes will be discussed in Section 5 below.

When studying a given process one also needs to take into account contributions from pion loops, which produce divergences that are eliminated by a renormalization of the low--energy constants of the effective lagrangian. We have computed this divergent part of the one--loop generating functional in the even intrinsic parity sector using standard heat--kernel techniques, and we have then performed the corresponding renormalization of the low--energy constants in the same dimensional renormalization scheme as described in \cite{GL1,GL2}.
Thus the low--energy constants display a  logarithmic scale dependence ($X(\mu )$ denotes
generically any of these renormalized low-energy constants)
\eq
X(\mu )= X(\mu ') + {{\Gamma_X}\over{(4\pi )^2}}\,\cdot\ln({\mu '}/\mu )\ .
\lbl{scale}
\en 
The full list of the resulting $\beta$--function coefficients $\Gamma_X$ is given in Table 1 below.\footnote{~These results have also been established independently by L.~Girlanda, private communication to M.K. and \cite{girl98}. The renormalization of the ${\cal L}_{(4,0)}^+$ counterterms has, of course, already been obtained before in \cite{GL1}.}

\begin{center}
\begin{tabular}{c|ccc|ccc|c} 
 $X$   & $F^2\cdot\Gamma_X$&$\qquad\qquad$&$X$   & $F^2\cdot\Gamma_X$& 
$\qquad\qquad$&$X$   & $\Gamma_X$\\ \cline{1-2} \cline{4-5}\cline{7-8}
$A$   & $3B^2$ & & $\xi^{(2)}$  & $4B$& &$l_1$    & ${1\over 3}$\\ \cline{4-5}
$Z^P$ & $-{3\over 2}B^2$ & &$\rho_1^{(2)}$ & $-4B(A+Z^P)$ & &$l_2$ &${2\over 3}$
\\ \cline{1-2}
$h_0$   & $0$ & &$\rho_2^{(2)}$ & $-4B(A-3Z^P)$ & &$l_5$ & $-{1\over 6}$\\ 
$h_1$  & $6B^2$ & &$\rho_3^{(2)}$ & $2B(A+3Z^P)$ & &$l_6$ & $-{1\over 3}$\\ 
\cline{7-8}
        &        & &$\rho_4^{(2)}$   & $2B(3A+Z^P)$& &$h_2$    & ${1\over 12}$\\
        &        & &$\rho_5^{(2)}$   & $4B(A-2Z^P)$&         &
\end{tabular}
\vskip 0.5 true cm
{\bf Table 1a:} {\it Scale dependence of the low energy constants
of ${\tilde{\cal L}}^{(2)}$, ${\tilde{\cal L}}^{(3)}$ and of ${{\cal L}}_{(4,0)}^{+}$.}
\end{center}

\begin{center}
\begin{tabular}{c|ccc|c} 
     $X$   & $F^2\cdot\Gamma_X$ & & $X$   & $F^2\cdot\Gamma_X$\\ 
\cline{1-2}\cline{4-5}
$a_1$      & $-2Z^P$    &$\qquad$ &$e_1$      & $-4A^2-22(Z^P)^2-20AZ^P$\\ 
$a_2$      & $-12Z^P$   &$\qquad$ &$e_2$      & $-4A^2-16(Z^P)^2-12AZ^P$\\ 
$a_3$      & $0$        &$\qquad$ &$e_3$     & $-12A^2-64(Z^P)^2-48AZ^P$\\  
\cline{1-2}\cline{4-5}
$b_1$      & $6(A+Z^P)$ &$\qquad$ &$f_1$      & $3A^2+15(Z^P)^2+12AZ^P$\\ 
$b_2$      & $-2(A+Z^P)$&$\qquad$ &$f_2$      & $2A^2+8(Z^P)^2+10AZ^P$\\ 
$b_3$      & $0$        &$\qquad$ &$f_3$      & $4A^2+24(Z^P)^2+20AZ^P$\\  \cline{1-2}
$c_1$      & $2(A+2Z^P)$&$\qquad$ &$f_4$      & $-6(Z^P+A)Z^P$\\ 
$c_2$      & $0$        &$\qquad$ &$f_5$      & $2A^2+8(Z^P)^2+10AZ^P$\\  \cline{4-5}
$c_3$      & $0$        &$\qquad$ &$h_4$      & $4A^2+8(Z^P)^2+8AZ^P$\\ 
$c_4$      & $2A$       &$\qquad$ &$h_5$      & $-4A^2-32(Z^P)^2-28AZ^P$\\  
$c_5$      & $0$        &$\qquad$ &$h_6$      & $4A^2+6(Z^P)^2+8AZ^P$\\  \cline{1-2}
$h_3$      & $0$        &$\qquad\qquad$ &$h_7$      & $-14(Z^P)^2$\\ 
\end{tabular}
\vskip 0.5 true cm
{\bf Table 1b:} {\it Scale dependence of the low energy constants
of ${\cal L}_{(2,2)}$ and of ${\cal L}_{(0,4)}$.}
\end{center}

Notice that at order ${\cal O}({\mbox p}^4)$, the low-energy constants of 
${\tilde{\cal L}}^{(2)}$ and
${\tilde{\cal L}}^{(3)}$ also need to be renormalized. The corresponding
counterterms, however, are of order ${\cal O}(B^2)$ and ${\cal O}(B)$,
respectively, and they are gathered in the  three last terms of Eq.  \rf{L4}~: 
in GChPT, renormalisation proceeds order by order in the
expansion in powers of $B$. Alternatively, one may think of Eqs.  \rf{L2} 
and
 \rf{L3} as standing for the combinations 
${\tilde{\cal L}}^{(2)} + B^2{\cal
L}'_{(0,2)}$ and ${\tilde{\cal L}}^{(3)} + B{\cal L}'_{(2,1)} + B{\cal
L}'_{(0,3)}$, respectively, with the corresponding low-energy constants
representing the renormalized, scale dependent, quantities. We shall adopt the latter point of view in the sequel. 

We have not worked out  the general structure of the divergent part of the 
one--loop generating functional in the odd intrinsic parity sector.
For SChPT, this has been done in Refs.~\cite{DonWyl,Issl,BijBraCor}.

\section{The One--Loop Amplitudes}
\setcounter{equation}{0}

Having constructed the effective lagrangian in the preceding section, the computation 
of the amplitudes ${\cal A}^N$ and ${\cal A}^C$ at next--to--leading order is a straightforward exercise.
We begin with the one--loop expression of the neutral amplitude ${\cal A}^N$, whose structure at that level of the chiral expansion is given by \rf{Aneu}, with
\bea
{\cal A}^N_1(k,k';p_1,p_2,p_3)&=&{{{\cal A}^{\pi^0\to\gamma\gamma}
{\cal A}^{00;00}(p_{12}^2,p_{13}^2,p_{23}^2)}\over{s-M_\pi^2}}
\nonumber\\
&&
-\frac{e^2}{4 \pi^2 F_\pi^3}\bigg\{
\frac{1}{2}(\beta -1)+\frac{1}{2}\mhat t (\beta -3)
-\alpha M_{\pi}^2t'-4\mhat^2t''-\gamma_{00}
\nonumber\\
&&
+\frac{1}{F_\pi^2} (\lambda_1+2\lambda_2)(s-3M_\pi^2))
\nonumber\\
&&
+\frac{1}{F_\pi^2} {\bar J}(p_{12}^2)\big[p_{12}^2-M_\pi^2+
{1\over 3}(\alpha-1)M_\pi^2\big]
\nonumber\\
&&
+\frac{1}{F_\pi^2} {\bar J}(p_{13}^2)\big[p_{13}^2-M_\pi^2+
{1\over 3}(\alpha-1)M_\pi^2\big]
\nonumber\\
&&
+\frac{1}{F_\pi^2} {\bar J}(p_{23}^2)\big[p_{23}^2-M_\pi^2+
{1\over 3}(\alpha-1)M_\pi^2\big]
\nonumber\\
&&
+\frac{2}{F_\pi^2}
\big[{\bar R}(p_{12}^2,k\cdot p_{12})+{\bar R}(p_{12}^2,k'\cdot p_{12})\big]
\big[p_{12}^2-M_\pi^2+\frac{1}{3}(\alpha-1)M_\pi^2\big]
\nonumber\\
&&
+\frac{2}{F_\pi^2}
\big[{\bar R}(p_{13}^2,k\cdot p_{13})+{\bar R}(p_{13}^2,k'\cdot p_{13})\big]
\big[p_{13}^2-M_\pi^2+\frac{1}{3}(\alpha-1)M_\pi^2\big]
\nonumber\\
&&
+\frac{2}{F_\pi^2}
\big[{\bar R}(p_{23}^2,k\cdot p_{23})+{\bar R}(p_{23}^2,k'\cdot p_{23})\big]
\big[p_{23}^2-M_\pi^2+\frac{1}{3}(\alpha-1)M_\pi^2\big]
\bigg\}\ ,
\nonumber\\
&&
\lbl{A1N}
\ena
and
\eq
{\cal A}^N_2(k,k';p_1,p_2,p_3)=
\frac{e^2}{2\pi^2 F_\pi^5}{\bar R}(p_{12}^2,k\cdot p_{12})
\big[p_{12}^2-M_\pi^2+{1\over 3}(\alpha -1)M_\pi^2\big]\ .\lbl{A2N}
\en
This second amplitude, which was absent at tree level, is entirely generated by the pion loops.
The numerator of the contribution which develops a pole at $s=M_\pi^2$ in the expression \rf{A1N} of ${\cal A}^N_1$ is given by the product of ${\cal A}^{\pi^0\to\gamma\gamma}$, which is related to the on-shell amplitude of the $\pi^0\to\gamma\gamma$ decay through ${\cal A}(\pi^0\to\gamma\gamma)=-t_1(k,\epsilon,k',\epsilon '){\cal A}^{\pi^0\to\gamma\gamma}$, times the amplitude ${\cal A}^{00;00}(p_{12}^2,p_{13}^2,p_{23}^2)$  of virtual $\pi^0-\pi^0$ scattering ($p_{12}^2+p_{13}^2+p_{23}^2=s+3M_\pi^2$). The expressions, at order 
${\cal O}({\mbox p}^6)$ and order ${\cal O}({\mbox p}^4)$, respectively, of these 
amplitudes read
\eq
{\cal A}^{\pi^0\to\gamma\gamma}=\frac{e^2}{4 \pi^2 F_\pi}
\big[1+\mhat t+ M_\pi^2t' +\mhat^2 t'' +2\mhat^2a_3\big]\ ,\lbl{ggpi0}
\en
and
\bea
{\cal A}^{00;00}(p_{12}^2,p_{13}^2,p_{23}^2) &=&
{{\alpha M_\pi^2}\over{F_\pi^2}}
+\frac{1}{F_\pi^4} (\lambda_1+2\lambda_2)\big[(p_{12}^2-2M_\pi^2)^2
+(p_{13}^2-2M_\pi^2)^2+(p_{23}^2-2M_\pi^2)^2\big]
\nonumber\\
&&+\frac{1}{F_\pi^4} {\bar J}(p_{12}^2)\bigg[(p_{12}^2-\frac{4}{3}M_\pi^2
+\frac{\alpha}{3}M_\pi^2)^2+\frac{\alpha^2}{2}M_\pi^4\bigg]
\nonumber\\
&&
+\frac{1}{F_\pi^4} {\bar J}(p_{13}^2)\bigg[(p_{13}^2-\frac{4}{3}M_\pi^2
+\frac{\alpha}{3}M_\pi^2)^2+\frac{\alpha^2}{2}M_\pi^4\bigg]
\nonumber\\
&&
+\frac{1}{F_\pi^4} {\bar J}(p_{23}^2)\bigg[(p_{23}^2-\frac{4}{3}M_\pi^2
+\frac{\alpha}{3}M_\pi^2)^2+\frac{\alpha^2}{2}M_\pi^4\bigg]\ .\lbl{A0000}
\ena

The various parameters $\al$, $\beta$, $\lambda_{1,2}$ and $\gamma_{00}$ involved in 
the expressions \rf{A1N} and \rf{A0000} are given in terms of combinations of 
the low--energy constants of the effective lagrangian 
${\cal L}^{\mbox{\scriptsize eff}}$ 
and of chiral logarithms due to the pion loops. They read
\bea
{{F_\pi^2}\over{F^2}}M_\pi^2\alpha &=&
2\mhat B + 16\mhat^2A
\nonumber\\
&&
+\mhat^3(81\rho_1^{(2)}+\rho_2^{(2)}+164\rho_4^{(2)}+2\rho_5^{(2)})
-4M_\pi^2\mhat\xi^{(2)}
\nonumber\\
&&
+16\mhat^4(16e_1+e_2+32f_1+34f_2+2f_3+72f_4+6a_3A)
\nonumber\\
&&
-8M_\pi^2\mhat^2(2b_1-2b_2-a_3-4c_1)
\nonumber\\
&&
-{1\over{16\pi^2F_\pi^2}}
\big[ 2M_\pi^4 + 102\mhat^2M_\pi^2A+264\mhat^4A^2\big]
\ln{{M_\pi^2}\over{\mu^2}}
\nonumber\\
&&
-{1\over{16\pi^2F_\pi^2}}
\big[ {{M_\pi^4}\over 2} + 44\mhat^2M_\pi^2A+264\mhat^4A^2\big]\ ,
\lbl{alphaexp}
\ena

\bea
\beta &=& 1 + 2\mhat\xi^{(2)} - 4\mhat^2(\xi^{(2)})^2
+2\mhat^2(3a_2+2a_3+4b_1+2b_2+4c_1)
\nonumber\\
&&
-{{M_\pi^2}\over{48\pi^2F_\pi^2}}\bigg(\ln{{M_\pi^2}\over{\mu^2}} +1\bigg)
\big[ 6+5(\alpha -1)\big]\ ,
\nonumber\\
&&
\ena

\bea
\lambda_1 &=& {1\over{48\pi^2}}\big( {\bar l}_1 - {4\over 3} \big)\ ,
\nonumber\\
&&\lbl{l1l2}\\
\lambda_2 &=& {1\over{48\pi^2}}\big( {\bar l}_2 - {5\over 6} \big)\ ,
\nonumber
\ena

\bea
\gamma_{00}&=&\mhat^2(3a_2+2a_3+6b_2+12c_1)
\nonumber\\
&&-{{M_\pi^2}\over{32\pi^2F_\pi^2}}\bigg(\ln{{M_\pi^2}\over{\mu^2}}+1\bigg)
(\alpha-1)\ .\lbl{g00}
\ena
%
The parameters $t$, $t'$ and $t''$ contain the contributions from ${\cal L}_{(4,1)}^-$, ${\cal L}_{(6,0)}^-$ and ${\cal L}_{(4,2)}^-$, respectively. In 
particular, from the formulae \rf{L41-} and \rf{L60-} we derive
\bea
t &=& {32\over 3}A_4\ ,
\nonumber\\
t' &=& -{8\over 3}(A_2-2A_3)\ .\lbl{ttprime}
\ena
Finally, ${\bar J}(s)$ denotes the Chew--Mandelstam function~\cite{chew}, the 
usual scalar two--point loop integral subtracted at $s=0$ (for its expression, 
see Ref.~\cite{GL1}), whereas ${\bar R}(p^2,k\cdot p)$, which is related to 
the three--point scalar loop function, is given, for $k^2=0$, by
\eq 
{\bar R}(p^2,k\cdot p)={\bar C}(p^2, k\cdot p)-
{{(k-p)^2}\over{4(k\cdot p)}}\big[{{\bar J}((k-p)^2)-\bar J}(p^2)\big]
+{1\over{32\pi^2}}\ ,
\en
with ($\sigma = \sqrt{1-4M_\pi^2/p^2}$, $\sigma' = \sqrt{1-4M_\pi^2/(k-p)^2}$)
\eq
16\pi^2{\bar C}(p^2, k\cdot p) = {{M_\pi^2}\over{4(k\cdot p)}}\big[
\ln^2\big({{\sigma-1}\over{\sigma+1}}\big) -
\ln^2\big({{\sigma'-1}\over{\sigma'+1}}\big)\big]\ .
\en 

For the charged amplitude ${\cal A}^C$, the general structure is more involved, see Eq.~\rf{decomp}, and at one loop we obtain

\bea
{\cal A}^C_1(k,k';p_0,p_+,p_-)&=&
{{\cal A}^{\pi^0\to\gamma\gamma}
{\cal A}^{00;+-}(p_{+-}^2,p_{+0}^2,p_{-0}^2)\over{s-M_\pi^2}}
\nonumber\\
&&+\frac{e^2}{4\pi^2F_\pi^3}\bigg\{
\frac{1}{2}(\beta +1)+\frac{1}{2}\mhat t(\beta -1)
-\frac{1}{3}(\alpha -1)M_\pi^2t'\nonumber\\
&&+
\gamma_{+-}+\frac{2}{3}\gamma_{+-}'\nonumber\\
&&+
{{\lambda_1}\over{F_\pi^2}}(p_{+-}^2-2M_\pi^2)
+
{1\over{F_\pi^2}}(\lambda_2-{1\over{288\pi^2}})
(p_{+0}^2+p_{-0}^2-4M_\pi^2)
\nonumber\\
&&
+{1\over{6F_\pi^2}} \bar{J}(p_{+-}^2)\big[3p_{+-}^2+4(\alpha -1)M_\pi^2\big]
\nonumber\\
&&
+{1\over{12F_\pi^2}} \bar{J}(p_{+0}^2)\big[5p_{+0}^2-14M_\pi^2-2(\alpha -1)M_\pi^2\big]
\nonumber\\
&&
+{1\over{12F_\pi^2}} \bar{J}(p_{-0}^2)\big[5p_{-0}^2-14M_\pi^2-2(\alpha -1)M_\pi^2\big]\bigg\}\ ,
\ena

\bea
{\cal A}^C_2(k,k';p_0,p_+,p_-)&=&
\frac{e^2}{24\pi^2F_\pi^3}\bigg\{
6+6\gamma_{+-}'-\frac{1}{48\pi^2F_\pi^2}\big[(k'-p_0)^2+p_{+0}^2+p_{-0}^2\big]
\nonumber\\
&&
-{1\over{F_\pi^2}}{\bar J}(p_{+0}^2)\big[4M_\pi^2-p_{+0}^2\big]
-{1\over{F_\pi^2}}{\bar J}(p_{-0}^2)\big[4M_\pi^2-p_{-0}^2\big]
\nonumber\\
&&
-{1\over{F_\pi^2}}{\bar J}\big((k'-p_0)^2\big)\big[4M_\pi^2-(k'-p_0)^2\big]
\bigg\}\ ,
\ena

\eq
{\cal A}^C_3(k,k';p_0,p_+,p_-)=
-\frac{e^2}{4\pi^2F_\pi^5}
{\bar R}(p_{+-}^2;k\cdot p_{+-})\big[p_{+-}^2+{4\over 3}(\alpha -1)M_\pi^2
\big]\ ,
\en

\bea
{\cal A}^C_4(k,k';p_0,p_+,p_-)&=&
-\frac{e^2}{4\pi^2F_\pi^5}\bigg\{
\frac{1}{3}\bar{J}\big((k'-p_{-})^2\big) M^2_\pi 
(4\frac{k\cdot p_+}{k\cdot p_{+0}}-1)
\nonumber\\
&&
+\frac{1}{6}
\big[\bar{J}\big((k'-p_{-})^2\big)-\bar{J}(p_{+0}^2)\big]
\bigg[p_{+0}^2+2 M^2_\pi+\frac{1}{2}\frac{p_{+0}^2}{k\cdot p_{+0}}
(4M^2_\pi-p_{+0}^2)
\nonumber\\
&&\qquad
+\frac{k\cdot p_+}{k\cdot p_{+0}}(
4M^2_\pi-7p_{+0}^2+6 k \cdot p_{+0})
+2\frac{k\cdot p_+}{k\cdot p_{+0}} \frac{p_{+0}^2}{k\cdot p_{+0}}(p_{+0}^2-4M^2_\pi) \bigg]
\nonumber\\
&&
+{\bar R}(p_{+0}^2,k \cdot p_{+0})\big[-M^2_\pi+2 k\cdot p_+ 
(\frac{p_{+0}^2}{k\cdot p_{+0}}-1)-\frac{1}{3}(\alpha-1)M^2_\pi\big]
\nonumber\\
&&
+\frac{1}{24\pi^2} k\cdot p_+ (1-\frac{p_{+0}^2}{k\cdot p_{+0}})
+ \frac{1}{96 \pi^2} p_{+0}^2\bigg\}\ ,
\ena

\eq
{\cal A}^C_5(k,k';p_0,p_+,p_-)=
-\frac{e^2}{24\pi^2F_\pi^5}\bigg\{
{\bar J}\big((k'-p_{-})^2\big)\big[4M^2_\pi-(k'-p_{-})^2\big] 
-{\bar J}(p_{+0}^2)(4M^2_\pi-p_{+0}^2)
-\frac{1}{24\pi^2}k\cdot p_{+0}\bigg\}\ ,
\en 

\bea
{\cal A}^C_6(k,k';p_0,p_+,p_-)&=&
-\frac{e^2}{4\pi^2F_\pi^5}\bigg\{
\frac{1}{6}
\big[\bar{J}\big((k'-p_{-})^2\big)-\bar{J}(p_{+0}^2)\big]
\bigg[p_{+0}^2-4M^2_\pi -6 k\cdot p_+ 
\nonumber\\
&&\qquad+4 \frac{k\cdot p_+}{k\cdot p_{+0}}
(p_{+0}^2- M^2_\pi)
+\frac{k\cdot p_+}{(k\cdot p_{+0})^2}p_{+0}^2
(4M^2_\pi-p_{+0}^2)\bigg]
\nonumber\\
&&
-\frac{1}{3}\bar{J}\big((k'-p_{-})^2\big)\big[
k\cdot p_0+2 M^2_\pi\frac{k\cdot p_+}{k\cdot p_{+0}}\big]
\nonumber\\
&&
+{\bar R}\big(p_{+0}^2,(k-p_{+0})^2\big) (k \cdot p_+) \big[2-\frac{p_{+0}^2}
{k\cdot p_{+0}}\big]
\nonumber\\
&&
+\frac{1}{32 \pi^2} \frac{k\cdot p_+}{k\cdot p_{+0}}(k'-p_-)^2
+\frac{1}{96\pi^2}\bigg(\frac{4}{3} k\cdot p_+- p_{+0}^2 \frac{k\cdot p_+}
{k\cdot p_{+0}}\bigg)
\nonumber\\
&&
+\frac{1}{144 \pi^2} k\cdot p_{+0}\bigg\}\ .
\ena

The amplitude $A_1^C$ again contains a contribution with a pole at $s=M_\pi^2$, which is given by the product of the ${\cal O}({\mbox p}^6)$ $\pi^0\to\gamma\gamma$ amplitude ${\cal A}^{\pi^0\to\gamma\gamma}$ times the (off--shell) $\pi^0\pi^0\to\pi^+\pi^-$ amplitude ${\cal A}^{00;+-}$, with

\bea
-{\cal A}^{00;+-}(p_{+-}^2,p_{+0}^2,p_{-0}^2)
&&=
\ \frac{\beta}{F_\pi^2}(p_{+-}^2
-\frac{4}{3} M_\pi^2)+\frac{1}{3F_\pi^2}\alpha M_\pi^2\nonumber\\
&&+{{\lambda_1}\over{F_\pi^4}}(p_{+-}^2-2M_\pi^2)^2
  +{{\lambda_2}\over{F_\pi^4}}[(p_{+0}^2-2M_\pi^2)^2+(p_{0-}^2-2M_\pi^2)^2]
\nonumber\\
&&+{1\over{6F_\pi^4}} \bar{J}(p_{+-}^2)\bigg[4(p_{+-}^2-\frac{4}{3}M_\pi^2
+\frac{5}{6}\alpha M_\pi^2)^2-(p_{+-}^2-\frac{4}{3}M_\pi^2-\frac{2}{3}
\alpha M_\pi^2)^2 \bigg]
\nonumber\\
&&+{1\over{12F_\pi^4}} \bar{J}(p_{+0}^2)\bigg[3(p_{+0}^2-\frac{4}{3}M_\pi^2
-\frac{2}{3}\alpha M_\pi^2)^2+(p_{+-}^2-p_{-0}^2)(p_{+0}^2-4M_\pi^2) 
\bigg]
\nonumber\\
&&+{1\over{12F_\pi^4}} \bar{J}(p_{-0}^2)\bigg[3(p_{-0}^2-\frac{4}{3}M_\pi^2
-\frac{2}{3}\alpha M_\pi^2)^2+(p_{+-}^2-p_{+0}^2)(p_{-0}^2-4M_\pi^2) 
\bigg]\ ,
\nonumber\\
\ena
where $p_{+-}^2+p_{+0}^2+p_{-0}^2=s+3M_\pi^2$.

The remaining parameters $\gamma_{+-}$ and $\gamma_{+-}'$ which appear in the amplitudes ${\cal A}_1^C$, ... ${\cal A}_6^C$ contain the contributions from the low--energy constants and chiral logarithms~: 

\bea
\gamma_{+-}&=& -\mhat^2(a_2+2b_2+4c_1)
\nonumber\\
&&+{{M_\pi^2}\over{96\pi^2F_\pi^2}}\ln{{M_\pi^2}\over{\mu^2}}
(\alpha-1)
+{{M_\pi^2}\over{96\pi^2F_\pi^2}}
(\alpha-\frac{7}{3})+\mhat^2\delta\gamma_{+-}\ ,\lbl{g}
\ena

\bea
\gamma_{+-}' &=&
8\mhat(2A_{12}-A_{13}) + 8M_\pi^2(A_7-A_8)+6\mhat^2a_3\nonumber\\
&&
-{{M_\pi^2}\over{32\pi^2F_\pi^2}}\ln{{M_\pi^2}\over{\mu^2}}
+\mhat^2\delta\gamma_{+-}'\ .\lbl{g'}
\ena

With the expressions given above, it is straightforward to check that the three conditions listed before Eq. \rf{iso1} as well as Eq. \rf{iso2} are satisfied, which provides a non--trivial check of our calculation. The isospin relation \rf{iso1} implies that the condition
\eq
2 + 3\gamma_{+-}+\gamma_{00}+{{M_\pi^2}\over{24\pi^2F_\pi^2}}
=\mhat^2( 2a_3 - 3t'')
\en
must hold. This requires that the contributions of the ${\cal L}_{(4,2)}^-$ counterterms to $\gamma_{+-}$, which we have denoted as $\mhat^2\delta\gamma_{+-}$ 
in the expression \rf{g}, have to satisfy
\eq
3t''+3\delta\gamma_{+-}=0\ .\lbl{isocond}
\en
The contributions of the ${\cal L}_{(4,2)}^-$ counterterms to $\gamma_{+-}'$, 
$\mhat^2\delta\gamma_{+-}'$ in the expression \rf{g'}, are not constrained 
by  isospin symmetry.

Upon using the information provided by Table~1, it is straightforward to check that $\al$, $\beta$ and $\gamma_{00}$ are scale independent by themselves (the parameters $\lambda_1$ and $\lambda_2$ were directly expressed in terms of the scale independent quantities ${\bar l}_1$ and ${\bar l}_2$ defined in Ref.~\cite{GL1}). In order that the amplitudes ${\cal A}^{\pi^0\to\gamma\gamma}$, 
${\cal A}^N$ and ${\cal A}^C$ be independent of the subtraction scale $\mu$, the parameters $t$, $t'$, $t''$, $\gamma_{+-}$ and $\gamma_{+-}'$ must be 
separately scale independent. This requires that $\delta\gamma_{+-}$ and 
$\delta\gamma_{+-}'$ themselves are scale independent. Since we have not worked out the 
structure of 
the one--loop divergences of the generalized generating functional in the odd 
intrinsic parity sector, we could not perform these checks explicitly. 

>From the above formulae, one may infer the expressions of the amplitudes in the standard case \cite{pere96,perethesis}. Since only the result for the neutral amplitude ${\cal A}^N$  was  displayed explicitly in Ref.~\cite{pere96}, and the expressions of the amplitudes ${\cal A}^C_i$ in the charged channel are only available from the unpublished work \cite{perethesis}, we describe in some detail the necessary steps to obtain them. 
Their general structure is of course unchanged, the differences occur only in the expressions of the various combinations of low--energy constants that are involved. In particular, the contributions from ${\cal L}_{(0,3)}$, ${\cal L}_{(2,2)}$, ${\cal L}_{(0,4)}$, and ${\cal L}^-_{(4,2)}$ are relegated to higher orders. For the remaining constants, the correspondance with the usual SChPT notation is given as follows 
\bea
\hat m \xi^{(2)}_{\mbox{\scriptsize st}}
&=& \frac{1}{16\pi^2 F_\pi^2}(\bar l_4+ \ln \frac{M_\pi^2}
{\mu^2})\ , \nonumber \\
\alpha_{\mbox{\scriptsize st}}
&=& 1 + \frac{M_\pi^2}{32 \pi^2 F_\pi^2}(4 \bar l_4 -3 \bar l_3 -1)\ ,
\nonumber \\
\beta_{\mbox{\scriptsize st}}
&=& 1 + \frac{M_\pi^2}{8 \pi^2 F_\pi^2}( \bar l_4 -1)\ , \nonumber \\
\gamma_{00,{\mbox{\scriptsize st}}}&=& 0\ ,\nonumber  \\
\gamma_{+-,{\mbox{\scriptsize st}}}&=& 
- \frac{M_\pi^2}{72 \pi^2 F_\pi^2}\ , \nonumber \\
\gamma_{+-,{\mbox{\scriptsize st}}}'&=& 8M_\pi^2 (A_7-A_8+ 
\frac{2 A_{12}-A_{13}}{2B})- \frac{M_\pi^2}{32 \pi^2 F_\pi^2}
\ln 
\frac{M_\pi^2}{\mu^2}\ .
\lbl{Sval}
\ena
We have checked that upon substituting these expressions into the one--loop amplitudes ${\cal A}^N$ and ${\cal A}^C$ given above, we recover the results of the standard case, up to the contributions from the counterterms 
contained in $t$ and $t'$, which were not included in 
Refs.~\cite{pere96,perethesis}.

\section{Counterterm Estimates and Numerical Results}
\label{COUNTER}
\setcounter{equation}{0}

In order to make numerical estimates for the cross sections based on the ${\cal O}({\mbox p}^6)$ amplitudes, we first need to fix or estimate the values of 
the various counterterms involved. 

\noindent
~~{\bf i)}~$\underline{ \lambda_1,\ \lambda_2}$~:
At order ${\cal O}({\mbox p}^4)$, these parameters are related to ${\bar l}_1$ and ${\bar l}_2$ through Eq.~\rf{l1l2}.
The values of these low--energy constants in the standard case have been the subject of numerous studies in the literature 
\cite{GL1,l1l2Kl4,l1l2coll,BCEGS,girlanda97}. A first determination at order ${\cal O}({\mbox p}^4)$ was given in Ref.~\cite{GL1}, using information from the D--wave $\pi$--$\pi$ scattering lengths. The corresponding values, taken from a recent numerical re--analysis~\cite{l1l2G}, are 
\eq
{\bar l}_{1,{\mbox{\scriptsize GL}}} = -2.15\pm 4.30\ ,
\ \ {\bar l}_{2,{\mbox{\scriptsize GL}}} = 5.84\pm 1.72\ ,
\en 
leading to 
\eq
\lambda_{1,{\mbox{\scriptsize GL}}} = (-7.35\pm 9.06)\times 10^{-3}\ ,
\ \ \lambda_{2,{\mbox{\scriptsize GL}}} = (10.57\pm 3.63)\times 10^{-3}\ .\lbl{l1l2G}
\en
The parameters $\lambda_1$ and $\lambda_2$ can also be determined directly,
{\it via} a set of rapidly convergent sum--rules \cite{pipipaper2}, 
from the knowledge, at order two loops, of the $\pi$--$\pi$ scattering amplitude $A(s\vert t,u)$ in GChPT \cite{pipipaper1}, and from medium energy data on $\pi$--$\pi$ phase shifts. The values obtained this way correspond to a determination at order ${\cal O}({\mbox p}^6)$. 
They depend only very weakly on the values of the parameters $\alpha$ and $\beta$ when the latter are varied within the ranges specified below, and read
\eq
\lambda_{1} = (-6.1\pm 2.2)\times 10^{-3}\ ,
\ \ \lambda_{2} = (9.6\pm 0.5)\times 10^{-3}\ .\lbl{l1l2K}
\en
These values are compatible with those given in eq.~\rf{l1l2G}, but are affected by much smaller error bars. The analysis may even be refined in the standard case, using the information on the SChPT two--loop $\pi$--$\pi$ amplitude obtained in Ref.~\cite{BCEGS}, leading to the following values~\cite{girlanda97},
\eq
\lambda_{1,{\mbox{\scriptsize st}}} = (-5.7\pm 2.2)\times 10^{-3}\ ,
\ \ \lambda_{2,{\mbox{\scriptsize st}}} = (9.3\pm 0.5)\times 10^{-3}\ .
\lbl{l1l2Kst}
\en

\noindent
~~{\bf ii)}~$\underline{\alpha,\ \beta}$~:
At leading order, $\alpha$ is directly correlated to the size of the condensate, see Eq.~\rf{alphalead}. As such, the value of $\alpha$ is not predicted by GChPT, but remains a free parameter, that can {\it a priori} be varied in the range $1\lapprox\alpha\lapprox 4$. At order ${\cal O}({\mbox p}^4)$, the relationship between $\alpha$ and the ratio $2\mhat B/M_{\pi}^2$ becomes more complicated, as shown in Eq.~\rf{alphaexp}. The corresponding next--to--leading and 
next--to--next--to--leading corrections have been estimated in Ref.~\cite{pipipaper1} (see in particular the figures 8 and 10 in that reference). Notice that the analysis of 
Ref.~\cite{pipipaper1} was done within the framework of 
$SU(3)_{\mbox{\scriptsize L}}\times SU(3)_{\mbox{\scriptsize R}}$ chiral 
perturbation theory. Working with only two light flavours as in the present 
paper might further reduce the uncertainties in the correspondance between 
the value of $\alpha$ and the size of the condensate, due to the lower number 
of unknown counterterms involved, and due to the absence of large 
contributions from the chiral logarithms induced by the kaon loops 
\cite{girl98}. However, the results of Ref.~\cite{pipipaper1} are sufficient for our present purposes, and we shall not pursue that matter further. Once $\alpha$ is given, the parameter $\beta$ is also constrained by low--energy $\pi$--$\pi$ data. The correlation between $\alpha$ and $\beta$, which results from the Morgan--Shaw universal curve~\cite{morganshaw}, has also been studied beyond leading order in \cite{pipipaper1}, and is summarized in Fig.~6 of that reference. For the subsequent numerical analyses, we shall take the values given in Table 2 below. The values given in the second line of this table correspond to the one--loop values $\alpha_{\mbox{\scriptsize st}}$ and 
$\beta_{\mbox{\scriptsize st}}$ of 
the standard case, which follow from the expressions given in Eq.~\rf{Sval}, 
and from the values ${\bar l}_3 = 2.9\pm 2.4$ \cite{GL1}, ${\bar l}_4 = 
4.4\pm 0.3$ \cite{pere98}. Two points are worth being remembered. The first 
is that, independently of the value of $\alpha$, $\beta$ stays close to 
unity. The second point is that the standard case allows to make a very 
precise prediction for the value of $\alpha$ at order ${\cal O}({\mbox p}^4)$,
 {\it viz.} $\alpha_{\mbox{\scriptsize st}} 
= 1.06\pm 0.06$. Furthermore, this value is barely affected by the 
corrections at order 
${\cal O}({\mbox p}^6)$~: The analysis of 
Ref.~\cite{girlanda97}, based on the results of \cite{BCEGS}, gives 
$\alpha_{\mbox{\scriptsize st}}=1.07\pm 0.01$ and 
$\beta_{\mbox{\scriptsize st}}=1.105\pm 0.015$ 
at next--to--next--to--leading order. Therefore, any significant deviation of 
the value of $\alpha$ from unity would provide evidence for a departure 
from the standard scenario of chiral symmetry breaking with a strong 
condensate.

\noindent
~~{\bf iii)}~$\underline{ t,\ t',\ t''}$~:
The constants $t$, $t'$ and $t''$ appear in the expression of the $\pi^0\to\gamma\gamma$ amplitude \rf{ggpi0}. The uncertainty on the experimental value of the decay rate~\cite{PDG}, $\Gamma({\pi^0\to\gamma\gamma})=7.74\pm 0.56$ eV, only yields a very weak constraint on the combination that appears in ${\cal A}^{\pi^0\to\gamma\gamma}$, {\it viz.} 
$\mhat t +M_\pi^2 t' +\mhat^2 t'' = (0.0\pm 3.6)\times 10^{-2}$. 
This is comparable to the estimate one would obtain through naive dimensional analysis \cite{dimann}. In addition, as shown in Ref.~\cite{bachir95}, isospin breaking effects can be sizeable in ${\cal A}^{\pi^0\to\gamma\gamma}$. Further information may be obtained by making use of the sum--rules considered in Ref.~\cite{bachir95}. The corrections due to $t'$ were however not taken into 
account there, but the analysis of \cite{bachir95} is easily extended to the more general situation. In the generalized case, a similar set of sum--rules can be established, but they do not yield complete information on the three constants $t$, $t'$ and $t''$. We have summarized this analysis in the Appendix for the interested reader. Here, we only quote the values that we shall use in the sequel (the estimate for $t''$ follows from naive dimensional analysis),
\eq
\mhat t\,=\, (6\pm 12)\times 10^{-3}\, ,
\ \ M_\pi^2t'\,=\,(-3\pm 3)\times 10^{-3}\, ,
\ \ \mhat^2 t''\,\sim\,\pm 1\times 10^{-3}
\ .\lbl{genWZresult}
\en

\noindent
~~{\bf iv)}~$\underline{ \gamma_{00},\ \gamma_{+-},\ \gamma_{+-}'}$~: 
The main difficulty in obtaining numerical estimates for these parameters comes from the lack of knowledge, in the generalized case, on the contributions from the low--energy constants of ${\cal L}^+_{(2,2)}$.
In the spirit of Vector Meson Dominance (VMD), we have estimated the  contributions of vector mesons to the two amplitudes ${\cal A}^N$ and ${\cal A}^C$, as done in Ref.~\cite{pere96} for the standard case. This way, we find that $t$, $t'$ and $t''$ receive no contribution, whereas the contribution to 
$\gamma_{00}$, $\gamma_{+-}$ and $\gamma_{+-}'$ read
\eq
\gamma_{00}\vert_{\mbox{\scriptsize VMD}}\,=\,
\gamma_{+-}\vert_{\mbox{\scriptsize VMD}}\,=\,0\ ,
\ \ \gamma_{+-}'\vert_{\mbox{\scriptsize VMD}}\,=\,
-\frac{3}{4}\frac{M_\pi^2}{M_V^2}\,\sim\,
-0.024\ .\lbl{VMDresult}
\en
We take \rf{VMDresult} as the values of these constants  at the scale $\mu\sim M_V$=770 MeV, and estimate the error associated to the VMD approximation and to the presence of the low-energy constants from ${\cal L}_{(2,2)}^+$ in the expressions of $\gamma_{00}$ and of $\gamma_{+-}$ by varying the scale $\mu$ of the corresponding chiral logarithms between 500 MeV and 1 GeV. The resulting values for $\gamma_{00}$ and $\gamma_{+-}$ are shown in Table 2. In the case of $\gamma_{+-}'$, we obtain a constant value which, to a very good precision, is compatible with zero.

\begin{table*}[htbp]
\begin{center}
\begin{tabular}{|c|c|c|c|}
\hline
 $\alpha$ & $\beta$ & $\gamma_{00}\times 10^3$ & $\gamma_{+-}\times 10^3$ \\
\hline
$1.06\pm 0.06$ & $ 1.103 \pm 0.008$ & $0$ & $-3$ \\
\hline
1.5 & $1.06 \pm 0.06$ & $8\pm 3$ & $-5.7\pm 0.8$ \\ 
\hline
2 & $1.07 \pm 0.06$ & $16\pm 5$ & $-8.4\pm 1.6$ \\
\hline
2.5 & $1.08 \pm 0.06$ & $24\pm 8$ & $-11.0\pm 2.4$ \\ 
\hline
3 & $1.11 \pm 0.06$ & $32\pm 10$ & $-13.7\pm 3.2$ \\
\hline
\end{tabular}
\end{center}
\begin{center}
{\bf Table 2:}{\it Values of $\beta$, $\gamma_{00}$ and $\gamma_{+-}$  for different values of $\alpha$.}
\end{center}
\end{table*}

\indent

With the above inputs at hand, we may now consider a few numerical 
applications. In Fig.~2, we have plotted the tree--level and 
one--loop cross sections for the charged channel, 
$\sigma^C_{\mbox{\scriptsize tree}}(s,\alpha)$
and $\sigma^C(s,\alpha)$,  
in the threshold region, $3 M_\pi \le \sqrt{s} \le 0.5$~GeV, 
where we expect the one--loop expression 
of the amplitude to be reliable, and for different values of $\alpha$. 
We have used $M_{\pi^\pm}
= M_{\pi^0}= 135$~MeV in the amplitude and the experimental values, 
as quoted in \cite{PDG}, in the phase space integrals.

Let us first concentrate on the standard case 
($\alpha\sim 1$) which has been discussed before in the literature. 
In the second half of the energy region that we have shown, the correction 
as compared to the tree level cross section amounts approximatively 
to a factor of two, if we consider the central values. 
This agrees with the unpublished result \cite{perethesis}, but is much 
less than previously found in \cite{pere96}. On the other hand, the error 
bars induced by the uncertainties attached to the various conterterm 
contributions 
that enter the ${\cal O}({\mbox p}^6)$ amplitude are important. 
A closer analysis 
(see also below) reveals that the main contribution comes from the uncertainty 
on the value of $\lambda_1$ given in \rf{l1l2Kst}. This shows also that the 
sensitivity of the cross section on the ${\cal O}({\mbox p}^6)$ counterterms 
becomes already sizeable even at such low energies. Coming now to the dependence with respect to $\alpha$, we see that 
here also the 
higher order corrections have a deep influence and upset the situation that 
prevailed at tree level.
For $\alpha\ge 2$ the 
one--loop cross sections are suppressed as compared to their tree--level 
values, and become even smaller than $\sigma^C(s,1)$ as the energy increases.
Unfortunately, the present theoretical 
error bars make it difficult to disentangle the different situations 
in practice from the knowledge of the total cross section alone.

\indent

\centerline{\psfig{figure=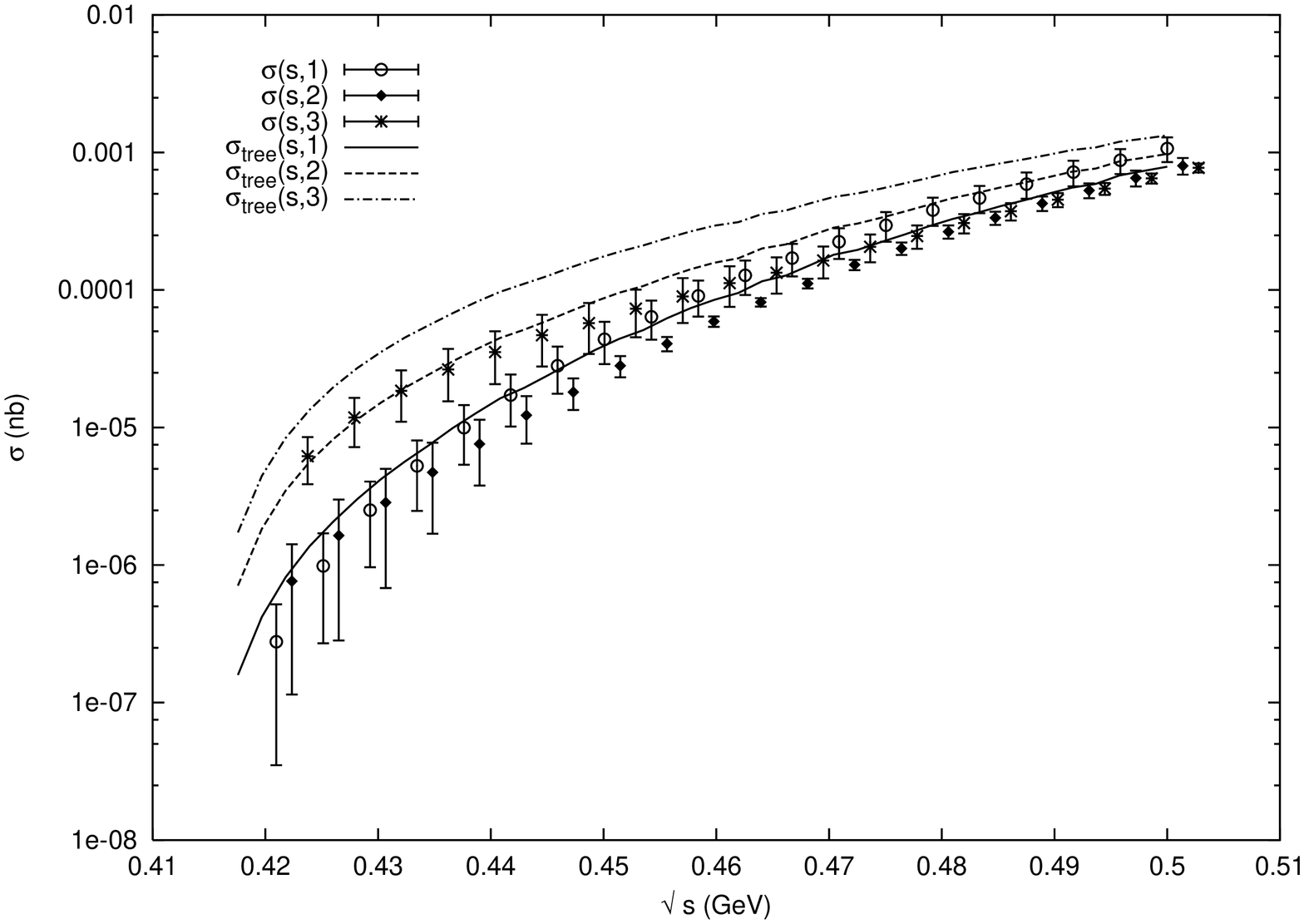,height=12.5cm,angle=-90}}
\noindent
{\bf Figure 2} : {\it The cross section $\sigma(s,\alpha)$
(in logarithmic scale) 
for $\gamma\gamma\to\pi^+\pi^-\pi^0$ 
as a function of the center of mass total energy, for three values of 
$\alpha$. Also shown are the corresponding curves for the tree level 
cross section {\rm $\sigma_{\mbox{\scriptsize tree}}(s,\alpha)$}.}

\indent

The cross sections $\sigma^N_{\mbox{\scriptsize tree}}(s,\alpha)$ 
and $\sigma^N(s,\alpha)$
in the neutral channel have been plotted in Fig.~3. Whereas 
$\sigma^N_{\mbox{\scriptsize tree}}(s,\alpha)\sim \alpha^2$, the 
corrections are seen to have an even more drastic influence on the behaviour 
of the cross 
section as a function of energy than in the charged channel. 
Unfortunately, as far as the dependence on $\alpha$ is concerned, 
the picture is again totally
blurred by the uncertainties, which, for the sake of clarity, 
we have not shown, but which are even 
more important than in the charged case.

In both channels, the origin of the large error bars 
is a consequence of the highly destructive 
interferences between the various amplitudes. 
In the charged case, this interference was already present at tree level, and 
is even accentuated by the loop effects. In the neutral case, the strong 
$\alpha$ dependence of the single amplitude that contributes at lowest order 
is, in a similar way, washed out by the interferences between the two 
one--loop amplitudes ${\cal A}_1^N$ and ${\cal A}_2^N$. 
In order to illustrate this point, we show, in Fig.~4, 
how the neutral cross section looks like if only the amplitude ${\cal A}_1^N$ 
is considered. The importance of the interference effects with the second 
amplitude ${\cal A}_2^N$ appears clearly upon comparing Figs.~3 and 4 (for the 
standard case, a similar observation was already made by the authors of 
Ref.~\cite{pere96}).
Unfortunately, we have not found a simple way to extract the contribution 
from 
${\cal A}_1^N$ only~: for photons with perpendicular polarizations, the two 
amplitudes contribute, and the destructive interference between them is 
again at work, while for photons with parallel polarizations, only the very 
small contribution from ${\cal A}_2^N$, which vanishes at tree level, is 
singled out. We have tried to investigate whether looking at more refined 
observables allows to reach better perspectives from this point of view. 

\indent

\centerline{\psfig{figure=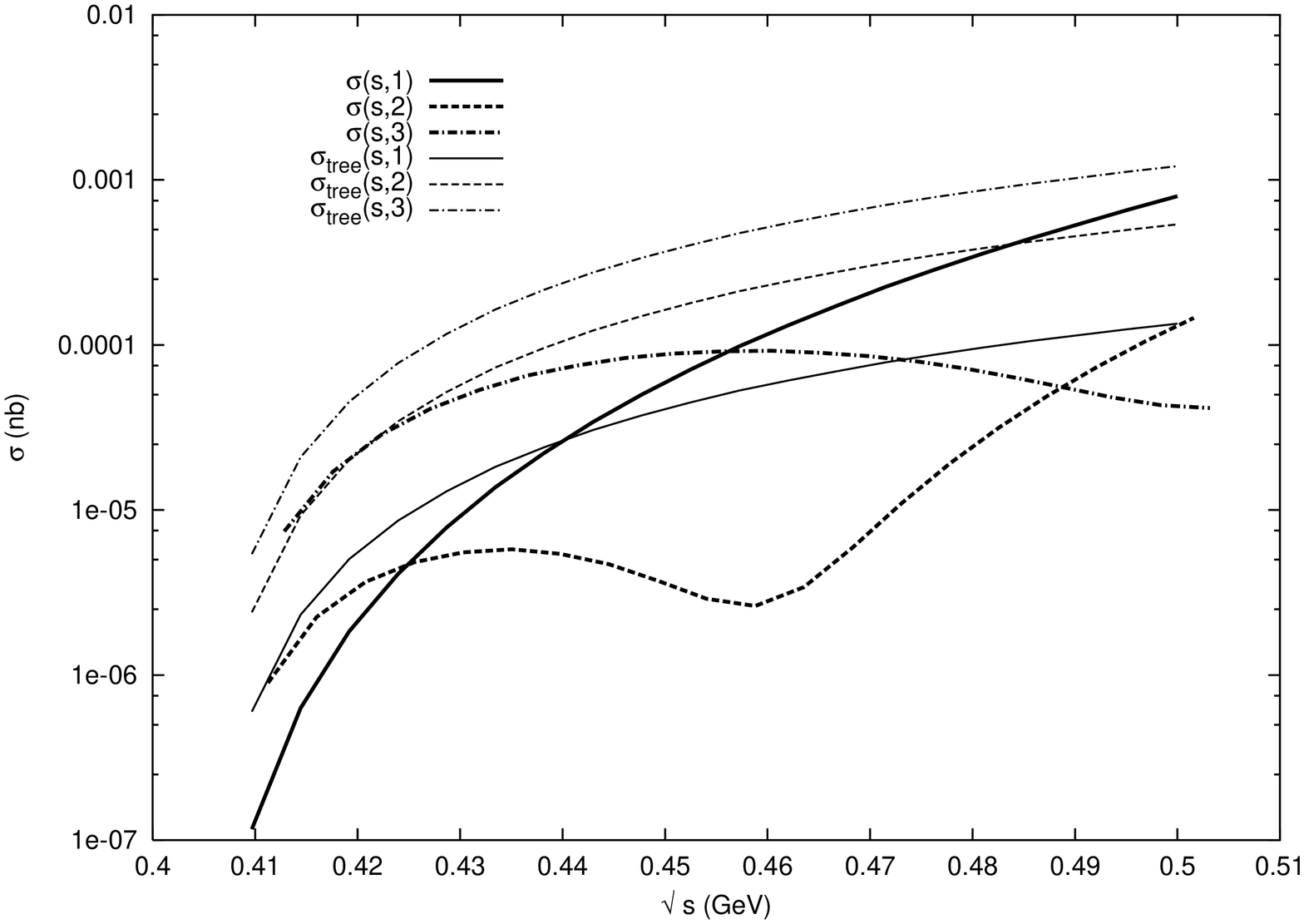,height=12.5cm,angle=-90}}
\noindent
{\bf Figure 3} : {\it The cross section $\sigma(s,\alpha)$
(in logarithmic scale) 
for $\gamma\gamma\to\pi^0\pi^0\pi^0$ 
as a function of the center of mass total energy, for three values of 
$\alpha$. Also shown are the corresponding curves for the tree--level 
cross section {\rm $\sigma_{\mbox{\scriptsize tree}}(s,\alpha)$}.}

\indent

\noindent
We have, for instance, looked at the invariant mass distribution of the two 
charged pions in the $\gamma\gamma\to\pi^+\pi^-\pi^0$ channel. The result is 
shown in Fig.~5. As one may observe, the error bars are much less important 
than for the total cross section, and the different values of $\alpha$ can be 
distinguished 
over a substantial portion of the energy range that has been considered.
 Actually, analyses of this type usually require sufficiently high 
statistics, which also represents a problem in the present case.

\indent

\centerline{\psfig{figure=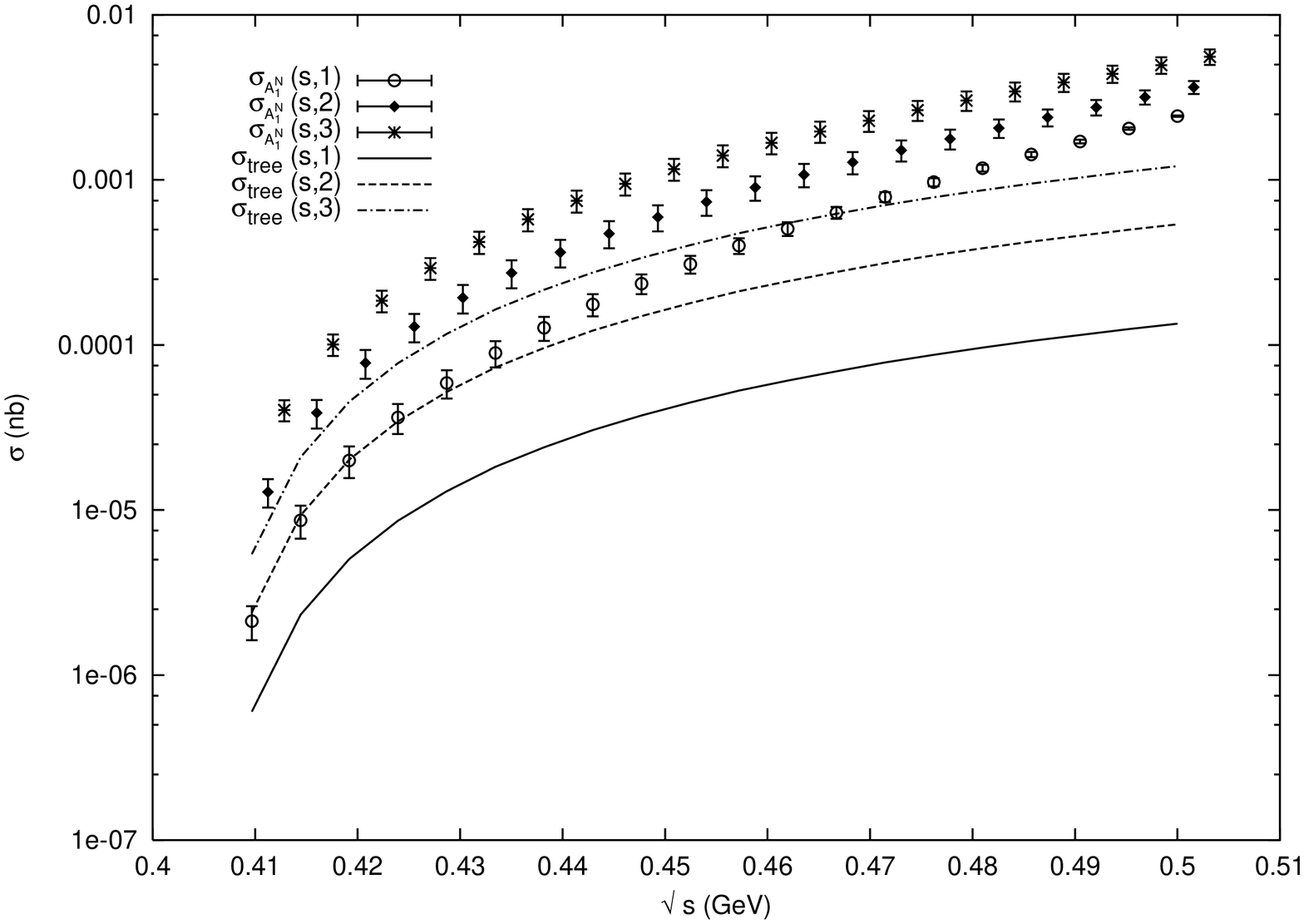,height=12.5cm,angle=-90}}
\noindent
{\bf Figure 4} : {\it The cross section $\sigma_{A_1^N}(s,\alpha)$ for 
$\gamma\gamma\to\pi^0\pi^0\pi^0$ obtained by taking into account the 
contribution from the amplitude $A_1^N$ alone, 
for different values of $\alpha$.}

\indent

Indeed, in both channels, the cross sections are very small at low energies, 
orders 
of magnitude below, for instance, the corresponding cross sections for the 
$\gamma\gamma\to\pi\pi$ processes. We have therefore also estimated the 
numbers of 
events that could be expected at an $e^+-e^-$ collider for the two--photon 
total invariant mass $\sqrt{s}$ below a maximal energy $E_{\mbox{\scriptsize 
max}}$.
As typical examples, we have considered two instances of symmetric $e^+-e^-$ colliders. The first 
case corresponds to the Daphne $\phi$--factory~\cite{daphne}, with a total beam energy 
of $E_{\mbox{\scriptsize beam}}\,=\,510$ MeV, and a nominal integrated luminosity 
of $5\times 10^6$ nb$^{-1}$ 
per year. The second case concerns a $\tau$--Charm Factory 
configuration~\cite{tauCF}, 
with a beam energy four times as large as for Daphne, and a design 
integrated luminosity of $10^7$ nb$^{-1}$ per year.
The numbers of events are obtained upon convoluting the above cross sections 
with the corresponding photon luminosities quoted in \cite{Lia}
(since we are not interested in resonant $\eta$ production,
whenever necessary we avoided the $\eta$ peak by applying a cut on
$m_{\gamma \gamma}$ such that only the events with $m_{\gamma \gamma}<
M_\eta -\Delta$ or $m_{\gamma \gamma} > M_\eta +\Delta$, with
$\Delta=20~$MeV are accepted),
\eq
{d L_{\gamma\gamma}\over d m_{\gamma\gamma}}=
{4\over m_{\gamma\gamma}}\big({\alpha\over \pi} \ln {E_{\mbox{\scriptsize beam}}\over m_e}\big)^2
\big[-(2+z^2)^2 \ln z -(1-z^2) (3+z^2) \big],
\en
where $m_{\gamma\gamma}=\sqrt{s}$ is the photon--photon center
of mass energy, $z=m_{\gamma\gamma}/2E_{\mbox{\scriptsize beam}}$, 
and $m_e$ is the electron mass.
For the luminosity quoted above, the expected total number of 
$\gamma\gamma\to\pi^+\pi^-\pi^0$ events per year 
at energies $\sqrt{s}\lapprox 0.6$~GeV is around $5\pm 1$ for Daphne 
(independently of $\alpha$) and even less in the neutral case. In the case of 
a $\tau$--Charm Factory, the total number of events becomes sizeable, and the 
results are shown, for the two modes, in the fourth column of Table~3,

\indent

\centerline{\psfig{figure=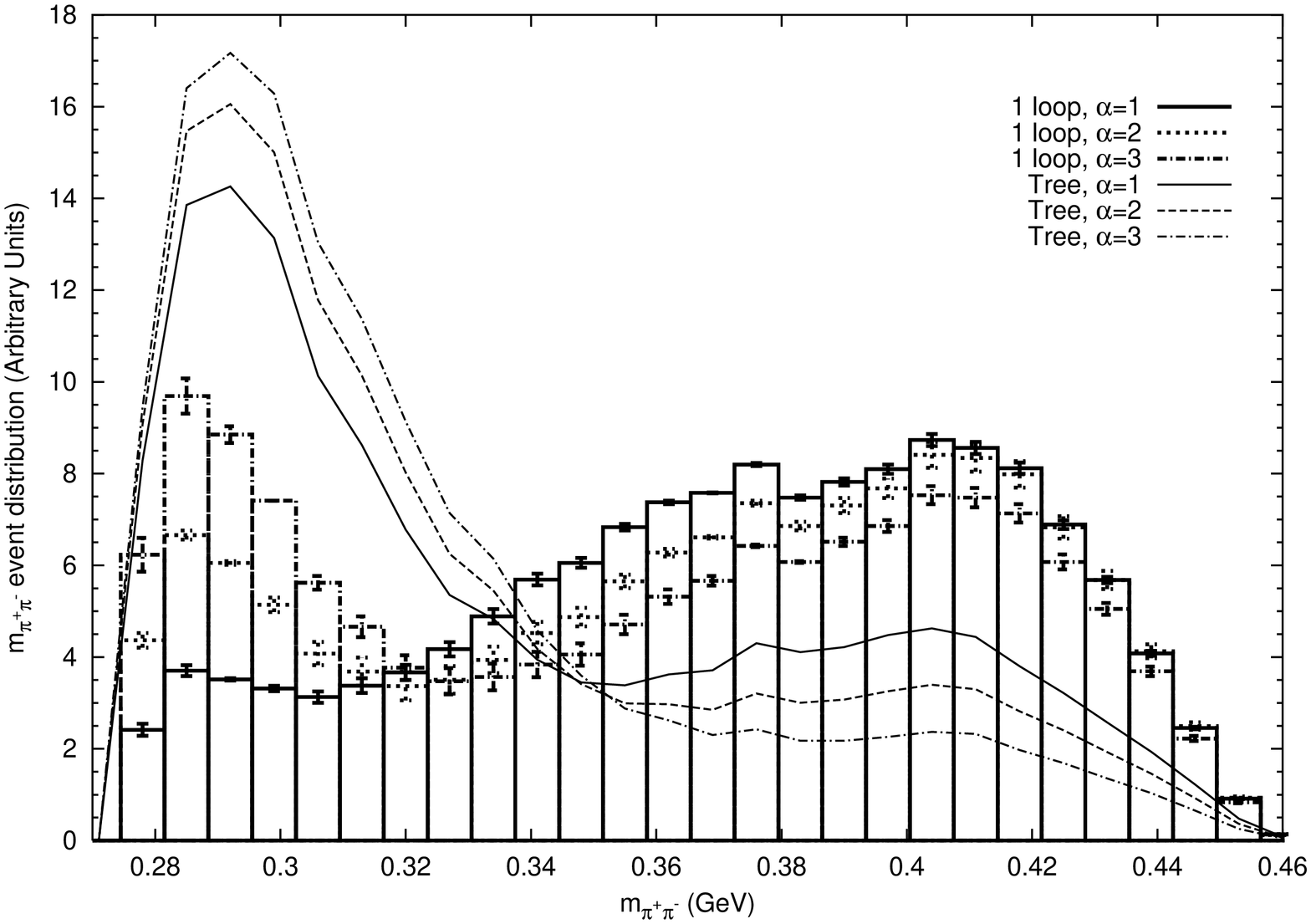,height=12.5cm,angle=-90}}
\noindent
{\bf Figure 5} : {\it The histogram of the distribution of the invariant mass $m_{\pi^+\pi^-}$ of the two final charged pions in the process $\gamma\gamma\to\pi^+\pi^-\pi^0$ for different values of $\alpha$ and $\sqrt{s} \le E^{max} = 0.6$~GeV. Also shown are the curves correponding to the same distribution, but from the lowest order amplitude alone.}

\indent

\noindent
for different choices 
of $E_{\mbox{\scriptsize max}}$ (second column) and for different values of 
$\alpha$ (third  
column). Also shown are the 
corresponding uncertainties $\Delta$(\#events) (fifth column), while the 
remaining entries show various sources of contributions to 
the total error $\Delta$(\#events). The latter was obtained 
upon adding the individual contributions in quadrature.
As mentioned above, the main source of error comes 
from the uncertainty on the value of $\lambda_1$. A sizeable but not necessarily drastic reduction of the 
latter would already allow to distinguish the standard case of a strong 
condensate ($\alpha\sim 1$) from situations where spontaneous breakdown 
of chiral symmetry would be triggered by a much weaker condensate ($\alpha
\gapprox 2$).

\indent

\noindent

\begin{table*}[t]
\begin{center}
\begin{tabular}{|c|c|c|c|c|c|c|c|c|}
\hline
Mode & $E_{\mbox{\scriptsize max}}({\mbox GeV})$ & $\alpha$ & \#events & $\Delta$(\#events) &
$\Delta\lambda_1$ & $\Delta\lambda_2$ & $\Delta\beta$ & $\Delta\mhat t$ 
\\ \hline
     &     & 1  & 16 & 4 & 3.3 & 1.3 & 0.3 & 1.3 \\
     &0.50 & 2  & 11 & 1 & 1   & 0.2 & 0.2 & 0.4 \\
     &     & 3  & 12 & 2 & 1   & 0.6 & 1.8 & 0.5 \\
\cline{2-9}
     &     & 1  & 56 &11 & 9.7 & 3.6 & 0.6 & 3.5 \\
$\pi^+\pi^-\pi^0$
     &0.55 & 2  & 40 & 5 & 4.5 & 1.3 & 0.1 & 1.8 \\
     &     & 3  & 38 & 3 & 0.2 & 0.6 & 3.1 & 0.1 \\
\cline{2-9}
     &     & 1  &327 &56 &50   & 18  & 1.8 & 14  \\
     &0.60 & 2  &273 &39 &35   & 11  & 2.7 & 10  \\
     &     & 3  &246 &26 &24   & 5.7 & 4.4 & 7.3 \\
\hline
\hline
     &     & 1  & 14 & 7   & 5.9 & 2.7 & 0.7 & 2.3 \\
     &0.50 & 2  & 1  & 2   & 1.2 & 0.6 & 1   & 0.5 \\
     &     & 3  & 5  & 4   & 2.6 & 1.2 & 2.6 & 1.3 \\
\cline{2-9}
     &     & 1  & 42 &18 &15.8 & 7.2 & 1.8 & 5.6 \\
$\pi^0\pi^0\pi^0$
     &0.55 & 2  & 8  & 9 & 5.9 & 2.7 & 4.8 & 2.1 \\
     &     & 3  & 6  & 3 & 1.7 & 0.8 & 2.0 & 1.0 \\
\cline{2-9}
     &     & 1  &211 &81 &71   & 32  & 6.3 & 19  \\
     &0.60 & 2  &91  &59 &45   & 20  & 28  & 12  \\
     &     & 3  &51  &34 &26   & 12  & 15  & 6.6 \\
\hline
\end{tabular}
\end{center}
{\bf Table 3:}~{\it Number of events for $\gamma\gamma\to\pi^+\pi^-\pi^0$ and 
for $\gamma\gamma\to\pi^0\pi^0\pi^0$ at a $\tau$--Charm Factory as a function 
of the maximal energy $E_{\mbox{\scriptsize max}}({\mbox GeV})$, and the 
principal sources of error.}
\end{table*}

\section{Summary and Conclusions}
\label{conclu}

In the present paper, the amplitudes of the  processes  
$\gamma\gamma\to \pi^0\pi^0\pi^0$ and 
$\gamma\gamma\to \pi^+\pi^-\pi^0$ have been computed in the framework of 
 $SU(2)_{\mbox{\scriptsize L}}\times SU(2)_{\mbox{\scriptsize R}}$ 
Generalized Chiral Perturbation Theory to  
${\cal O}({\mbox p}^6)$ precision. 
The corresponding generating functional has been 
constructed explicitly in Section 3, and the structure of its divergences in 
the sector of even intrinsic parity has been analysed. 
The resulting amplitudes, worked out in Section~4, 
satisfy the isospin relations that we have 
established in Section 2 (to the best of our knowledge, these relations have 
not been discussed previously in the literature). 
When restricted to the standard case, specified by the choice of 
parameters as indicated in Eq.~\rf{Sval}, we recover the results obtained by 
previous authors \cite{pere96,perethesis}, both in the neutral and in the 
charged case. Finally, we have estimated the counterterms that enter the 
one-loop amplitudes and we have performed some numerical analyses in Section 5.
We have, in particular, shown that the error bars associated to the 
counterterm estimates become important, especially in the neutral channel, 
as a consequence of highly destructive interference effects between 
the various amplitudes that build up the total cross sections.

We have also considered the possible detection of these processes at Daphne 
 and at a $\tau$--Charm Factory.
Unfortunately, and precisely because of these large interference effects, 
the expected number of events is rather discouraging in 
the first case.
Depending on the actual values of the counterterms and on $\alpha$, it is
hard to expect more than $\approx 5$ events per year with total invariant
mass lower than $500$~MeV.  The number of events increases substantially
when allowing larger invariant masses, but at the expense of 
working in an energy region where the ${\cal O}(p^6)$ 
expressions are probably less reliable, 
since higher order terms can become important.
The computation at ${\cal O}({\mbox p}^8)$ would allow a better control of 
the cross
sections at larger momenta, thus allowing a
substantial increase of the number of events already at Daphne.  
The required amount of work seems excessive, though, and would make sense only 
if conducted in parallel with a better determination of ${\bar l}_1$ (for instance, 
from a two--loop analysis of $K_{\ell 4}$ decays), by far the main source of theoretical uncertainties at present. We rather expect 
interesting and realistic prospects in 
this field to come from future machines, like the 
$\tau$--Charm Factory, which run at higher energy and higher luminosity.

\section*{Acknowledgments}
We are pleased to thank A. Bramon for discussions and for his interest and 
collaboration at the early stages of this work, and J.~Bijnens for discussions. One of us (M.K.) wishes to acknowledge clarifying correspondance and/or discussions with I.~Kogan and J.~Stern on the content of Ref.~\cite{KKS98}, and L.~Girlanda for discussions and for sharing information on unpublished work. He also thanks the Universitat Polit\`ecnica de Catalunya and the Universitat Aut\`onoma de Barcelona for their hospitality. Ll.~A. thanks the Institut de Physique Nucl\'eaire in Orsay for its hospitality. P.~T. thanks FEN department, where most of his 
contribution to
the present work was done. This work has been partially supported by the 
TMR Program, EC--Contract N. CT98--0169.
Ll.~A.  and P.~T.  received partial support from the CICYT research project 
AEN95--0815, while the work of J.~K. and of P.~T. 
was supported by the Schweizerischer Nationalfonds and 
by the Swedish Research Council (NFR), respectively. 

\indent

\indent

\noindent
{\Large{\bf Appendix}}

\renewcommand{\theequation}{A.\arabic{equation}}
\setcounter{equation}{0}

In this Appendix, we give a brief description of our analysis of the anomalous counterterms $A_2$, $A_3$, $A_4$ and $A_6$ which is based on the approach of Ref.~\cite{bachir95}. The starting point are the invariant amplitudes $\Pi_{VVP}(p^2,q^2,r^2)$ and $\Pi_{VVP}^0(p^2,q^2,r^2)$ of the vector--vector--pseudoscalar three--point correlation functions in the three flavour chiral limit (we take the definition given by Eqs.~(3) and (4) of \cite{bachir95}). At $p^2=q^2=0$, the loop contribution vanish, and one has (notice the absence of the pion pole in the second equality)
\bea
\Pi_{VVP}(0,0,r^2) &=& \frac{2B_0N_c}{16\pi^2r^2}\,+\,\frac{1}{4\pi^2}
\big[ 8A_4-16B_0(A_2-2A_3)\big]\ ,\nonumber\\
\Pi_{VVP}^0(0,0,r^2) &=& \frac{1}{4\pi^2}\big[ 8A_4 +24 A_6\big]\ .
\ena
In the chiral limit, the counterterms from ${\cal L}_{(4,2)}^-$ do not contribute, so that the above result holds both in SChPT and in GChPT. Although the contributions from the low--energy constants $A_2$ and $A_3$ from ${\cal L}_{6,0}^-$ were omitted in \cite{bachir95}, one may follow the same steps as described there (we have also kept the same notation), and end up with the following set of sum--rules~:
\eq
\frac{1}{4\pi^2}
\big[ 8A_4-16B_0(A_2-2A_3)\big] \,=\,
-\frac{B_0}{2M_V^2}\bigg\{ \frac{F_0^2}{M_V^2}\,+\,\frac{N_c}{4\pi^2}
\left(\frac{M_V}{M_P}\right)^2{\mbox{tan}}\Theta\frac{{\cal A}(\pi'\to\gamma\gamma)}{{\cal A}(\pi\to\gamma\gamma)}\bigg\}\ ,\lbl{bachir1}
\en
and
\eq
\frac{1}{4\pi^2}
\big[ 24A_6+16B_0(A_2-2A_3)\big] \,=\,-\frac{3}{8}
\frac{B_0G_{\eta'}}{M_{\eta '}^2}\sqrt{6}{\cal A}(\eta'\to\gamma\gamma)\ .
\lbl{bachir2}
\en
In addition, one needs to know the expression of the amplitude of the two--photon decay of the $\eta$, ${\cal A}(\eta\to\gamma\gamma )=-t_1(k,\epsilon,k',\epsilon'){\cal A}^{\eta\to\gamma\gamma}$, which, in analogy to the $\pi^0\to\gamma\gamma$ amplitude ${\cal A}^{\pi^0\to\gamma\gamma}$ of Eq.~\rf{ggpi0}, may be written as
\eq
{\cal A}^{\eta\to\gamma\gamma}\,=\,\frac{e^2}{4\sqrt{3} \pi^2 F_\pi}
\big[\frac{F_\pi}{F_\eta} +\frac{1}{3}\mhat (5-2r)t+
\frac{128}{3}\mhat (1-r)A_6+M_\eta^2t'+\mhat^2{\widetilde t}''(r)+\frac{2}{3}\mhat^2(1+2r^2)a_3\big]\ .\lbl{ggeta}
\en
In this last expression, $t$ and $t'$ are related to $A_2$, $A_3$ and $A_4$ according to Eq.~\rf{ttprime}, while $r$ stands for the quark mass ratio $m_s/\mhat$, and $\mhat^2{\widetilde t}''(r)$ denote the corrections coming from ${\cal L}_{(4,2)}^-$, which can be of order ${\cal O}(\mhat^2)$, ${\cal O}(\mhat m_s)$, and ${\cal O}(m_s^2)$.

In the standard case, the two last terms in Eq.~\rf{ggeta} would appear only at higher orders, and one may thus proceed as described in Ref.~\cite{bachir95}. At the order we are working, the quark mass ratio is then given as 
$r_{\mbox{\scriptsize st}} = r_2\equiv 2M_K^2/M_\pi^2-1\sim 25.9$~\cite{GL2}. 
Using the numerical values given in \cite{bachir95}, one obtains, from Eqs.~\rf{bachir1} and \rf{bachir2}, respectively,
\eq
\mhat A_{4,{\mbox{\scriptsize st}}}-M_\pi^2(A_2-2A_3)_{\mbox{\scriptsize st}} 
\,=\, (-5.9\pm 1.8)\times 10^{-4} 
\en
and
\eq
3\mhat A_{6,{\mbox{\scriptsize st}}}+M_\pi^2(A_2-2A_3)_{\mbox{\scriptsize st}} 
\,=\, (-2.1\pm 0.4)\times 10^{-3}\ . 
\en
Upon using the experimental rate for $\eta\to\gamma\gamma$~\cite{PDG}, we then determine the combination
\eq
M_\pi^2(A_2-2A_3)_{\mbox{\scriptsize st}} \,=\, (1\pm 1) \times 10^{-3} .\lbl{A2minus2A3std}
\en
Adding errors in quadrature, the previous results give
\eq
\mhat t_{\mbox{\scriptsize st}} + M_\pi^2t'_{\mbox{\scriptsize st}} = 
(1.7 \pm 8.2)\times 10^{-3}\ ,
\en
which is four times more accurate than the value obtained directly from the experimental rate of $\pi^0\to\gamma\gamma$. For the separate pieces, we obtain
\eq
\mhat t_{\mbox{\scriptsize st}} \,=\, (4.4\pm 10.8)\times 10^{-3}\, ,
\ \ M_\pi^2t'_{\mbox{\scriptsize st}} \,=\,(-2.7\pm 2.7)\times 10^{-3}
\ .\lbl{stdresult}
\en

In the generalized case, the analysis may, unfortunately, not be pursued quite that far. The main drawback are the corrections from ${\cal L}_{(4,2)}^-$, which in particular produce potentially large ${\cal O}(m_s^2)$ corrections to the $\eta\to\gamma\gamma$ decay amplitude, and on which the sum--rules \rf{bachir1} and \rf{bachir2} give no information. If one restricts the analysis to the order ${\cal O}({\mbox p}^5)$ precision, then the contributions from ${\cal L}_{(6,0)}^-$ are also absent, and the situation becomes even simpler than in the standard case, since ({\it cf.} Eq.~\rf{gencount}) the left--hand sides of the sum--rules \rf{bachir1} and \rf{bachir2} now only involve $A_4$ and $A_6$, respectively. Keeping in mind that $r$ is now a free parameter ($\alpha$ and $r$ are however related, see \cite{gchpt}), and taking the necessary inputs from \cite{bachir95}, the ${\cal O}({\mbox p}^5)$ determination  of $A_4$ and $A_6$ reads
\eq
\mhat A_4\,=\,-\frac{N_c}{32}\left(
\frac{\vert\lambda (r) M_S^2 - (1-\lambda (r))^2M_\pi^2\vert}{M_P^2-M_S^2}
\right)^{\frac{1}{2}}\frac{M_\pi}{M_P}\frac{{\cal A}(\pi'\to\gamma\gamma)}{{\cal A}(\pi\to\gamma\gamma)}\,+\,\cdots\ ,\lbl{A4gen}
\en
\eq
\mhat A_6\,=\,-\frac{3\pi^2}{32}\frac{M_\pi}{M_{\eta '}}\left(
\lambda (r) - \frac{\Delta_{\mbox{\scriptsize GMO}}}{(r-1)^2}\right)^{\frac{1}{2}}F_\pi
{\cal A}(\eta'\to\gamma\gamma)\,+\,\cdots\ ,\lbl{A6gen}
\en
where the ellipses stand for higher order corrections, $\lambda (r) = 2(r_2-r)/(r^2-1)$ and $\Delta_{\mbox{\scriptsize GMO}} \equiv (3M_\eta^2-4M_K^2+M_\pi^2)/M_\pi^2\sim -3.6$. For $r=r_2$, the above expressions give values compatible with the previous SChPT analysis. Furthermore, as $r$ decreases ({\it i.e} as $\alpha$ increases), the variation of $\mhat t$ stays within the bounds given in Eq.~\rf{stdresult}. On the other hand, in the extreme case of a vanishing condensate, the ${\cal L}_{(6,0)}^-$ contributions to the two sum--rules also disappear, and the expressions \rf{A4gen} and \rf{A6gen} become exact at order
${\cal O}({\mbox p}^6)$. Finally, if we use the two sum--rules in order to express $A_4$ and $A_6$ in terms of $A_2-2A_3$ in the expression \rf{ggeta}, we obtain an estimate of a combination of $A_2-A_3$, $\mhat^2{\widetilde t}''(r)$ and $\mhat^2a_3$, which is not very sensitive to the value of $r$ and compatible with the value \rf{A2minus2A3std} obtained in the standard case for $M_\pi^2(A_2-2A_3)$. Thus, within reasonable error bars, the values of $t$ and $t'$ can be taken independent of $r$. For the numerical analyses presented in the text, we have used
\eq
\mhat t\,=\, (6\pm 12)\times 10^{-3}\, ,
\ \ M_\pi^2t' \,=\,(-3\pm 3)\times 10^{-3}
\ .\lbl{genresult}
\en

\end{document}